%% file: ppnp_n.tex
\documentclass[twoside,12pt]{article}
\usepackage{epsfig}
\usepackage{amsmath,amssymb}

\newcommand{\be}{\begin{equation}}
\newcommand{\ee}{\end{equation}}
\newcommand{\bea}{\begin{eqnarray}}
\newcommand{\eea}{\end{eqnarray}}

\newcommand {\mbf}[1]{{\mathbf{#1}}}

\newcommand {\mcu}{\mathcal{U}}

\newcommand{\cm}{\mathrm{c\!\:\!.m\!\:\!.}}
\newcommand{\He}{{}^3\mathrm{He}}
\newcommand{\Hh}{{}^3\mathrm{H}}
\newcommand{\nH}{n\text{-}{}^3\mathrm{H}}
\newcommand{\pHe}{p\text{-}{}^3\mathrm{He}}
\newcommand{\pH}{p\text{-}{}^3\mathrm{H}}
\newcommand{\nHe}{n\text{-}{}^3\mathrm{He}}

\newcommand{\dd}{d\text{-}d}

\newcommand\MD{\mathfrak{D}}
\newcommand\BMD{\bar{\mathfrak{D}}}

\begin{document}

\title{ \vspace{1cm} Bound state techniques to solve the multiparticle
scattering problem}
\author{J.\ Carbonell,$^{1}$ A.\ Deltuva,$^2$ A. C. \ Fonseca,$^2$ R.\
Lazauskas,$^{3}$ \\
$^1$ Institut de Physique Nucl\'eaire Orsay, CNRS/IN2P3, F-91406 Orsay
Cedex, France.\\
$^2$Centro de F\'{\i}sica Nuclear da Universidade de Lisboa, P-1649-003
Lisboa, Portugal.\\
$^3$ Universit\'e de Strasbourg, IPHC, CNRS, 23 rue du Loess, 67037
Strasbourg, France.}
\maketitle

\begin{abstract}

Solution of the scattering problem turns to be very difficult task
both from the formal as well as from the computational point of
view. If the last two decades have witnessed decisive progress in
ab initio bound state calculations, rigorous solution of the
scattering problem remains limited to A$\leq$4 case. Therefore
there is a rising interest to apply bound-state-like methods to
handle non-relativistic scattering problems. In this article the
latest theoretical developments in this field are reviewed. Five
fully rigorous methods will be discussed, which address the
problem of nuclear collisions in full extent (including the
break-up problem) at the same time avoiding treatment of the
complicate boundary conditions or integral kernel singularities.
These new developments allows to use modern bound-state techniques
to advance significantly rigorous solution of the scattering
problem.

\end{abstract}


\include{Contrib_Antonio}
\include{Contrib_Jaume_cor}
\include{Contrib_Rimas}
\include{Contrib_Arnas_cor}
\section{Summary and Outlook \label{sec:out}}

The last two decades have witnessed a real revolution in the ab initio
treatment of the nuclear bound state problem based on nonrelativistic
Hamiltonians. Important steps have been taken in improving our
understanding of the interaction between nucleons, including
two- and three-nucleon forces~\cite{Epelbaum:2009,Machleidt:2011}.
Moreover, ab initio bound state methods that allow a solution of
the nuclear many-body problem without any uncontrolled
approximation have been evolved to treat systems with dozens of
nucleons~\cite{Pieper2002,Hagen:2010prc,Leidemann:2013ppnp,Barrett:2013ppnp}.

The ab initio treatment of nuclear collision problems has instead progressed at a modest pace.
Nevertheless, key steps have been also undertaken in this
direction. The first serious issue is related with the inclusion
of the Coulomb interaction for the collision above the
three-cluster breakup threshold. This problem has been overcome
both by momentum space as well as coordinate space methods, using
very different
techniques~\cite{kievsky:01a,deltuva:05a,deltuva:05d,LC11,DFL12}.
The next important step was to demonstrate the possibility of
exact solutions of multiparticle problems in $A>3$ systems above breakup threshold.
This has been also achieved  almost simultaneously by the
developments in momentum and coordinate space
frameworks~\cite{uzu:03a,deltuva:12c,La12}.

The next important challenge is to push forward calculations of collision problems to the level achieved by the modern bound
state methods, i.e. well beyond the $A=4$ system. This challenge is
well understood by the leading developers of bound-state
methods who started switching gears to the problem of nuclear
collisions. Yet, up to this moment, most developments involve approximate prescriptions of the
scattering
problem~\cite{Navratil:2010,Hagen:2012,Barrett:2013ppnp}. The main
difficulty of the collision problem is related to the complex
asymptotic form of the wave function in coordinate space, which
gives rise to a complex structure of singularities when
describing multiparticle dynamics using integral equations in
momentum space. Furthermore, the complexity of the wave function
asymptotic form (integral kernel singularities) quickly rises
once multiparticle breakup channels become open. Therefore the
importance of exact methods enabling to the treatment of the
multiparticle scattering problem by avoiding these formal as
well as technical difficulties is well understood and turns out to be
crucial for further developments.

The goal of this review is to demonstrate available tools to
solve the scattering problem exactly. We have overviewed five
different methods, namely: Lorentz Integral Transform, Continuum
discretization, Complex Scaling, Complex energy, and Momentum
lattice methods. All of these methods are able to handle
multiparticle scattering problems rigorously, both below and above breakup threshold and provide full information
about the respective integral as well as differential observables.

 The Lorentz Integral Transform method remains limited to the breakup of bound systems by an external perturbation; however this
method has been already implemented beyond the $A=4$ system and seems
to be the most easy to incorporate in many-body
calculations~\cite{Bacca:2013dma}.

The Continuum discretization method formally requires no
modifications in the existing bound state techniques to be
implemented to treat the scattering problem. One should only be
able to adjust the calculated positive energy eigenvalues to the energy of the collision. Difficulties in implementing
this method may however appear for systems with multiple
binary channels. One should find as many independent
solutions as there are channels, which may turn out into a
complex technical task.

The last difficulty is not present in the Complex Scaling or
Complex Energy methods. Implementation of these method is also
rather straightforward with any bound state method, once one
extends the available codes to handle complex arithmetics. This
may require not only simple modification of the arithmetical
operations used in the computer codes but also adjustment of the
linear algebra routines, which sometimes are only proper for real
Hermitian matrices. Nevertheless, very efficient iterative linear
algebra methods have been developed by the mathematicians and
software developers during the last decade~\cite{Saad:2003}, that
may handle algebraic problems for general complex matrices. An
efficient implementation of the  Complex Energy method requires,
however, special numerical techniques beyond the standard bound
state calculations.

Still one important step should be undertaken to handle these
methods in conjunction with the most-advanced bound state codes.
Solution of the scattering problem as well as calculation of the
observables requires construction of the incoming wave function.
This wave function is based on bound-state wave functions of
inherent clusters, and thus formally should be available.
Nevertheless composed cluster wave functions should be fully
consistent with the full Hamiltonian. This last issue may in fact be an
important handicap for some methods, since optimal variational
parameters used to calculate the N-body system may differ from the
ones applied in composed cluster calculations present in the incoming
wave. Likewise some similarity transformations used to make strong
the interaction easier to treat by bound state methods, like
Lee-Suzuki transform~\cite{Suzuki:1982}, are particle number
dependent and thus may not be appropriate for the scattering
problem. Nevertheless the similarity renormalization group
interactions which do not depend on the particle number are beeing
actively developed and have become very popular in bound-state
calculations.


We believe that the next decade will be  crucial for further
evolution in the many-body scattering problem. Moreover, experimental
installations of the next generation, based on radioactive ion
beams, are becoming available in the near future. These experiments require a
quantitatively better understanding and interpretation of the
nuclear collision
process~\cite{Kubono:2010,Thoennessen:2011,Tanihata:2013ppnp} given the unstable nature of the projectiles.
The recent developments we present in this manuscript may provide the
necessary tools to overcome serious difficulties in the solution
of the multiparticle scattering problem.

\subsubsection*{Acknowledgments} The authors thank G. Orlandini,
V.~N. Pomerantsev and V.~I. Kukulin for discussions and 
details of their calculations.




\bibliography{Bib_Arnas,Bib_Rimas}

\end{document}

%% file: Contrib_Antonio.tex
\section{Introduction}

Numerically exact solutions of the multi-particle scattering
problem have been, in the last 30 years, dominated by calculations
of the three-nucleon system
\cite{gloeckle:96a,kievsky:01a,deltuva:03a} where the early
development of formally exact theories \cite{faddeev:60a,alt:67a}
led to calculations that soon became numerically converged for a
number of realistic interaction models some of which
based on meson field theory
\cite{stoks:94a,wiringa:95a,machleidt:01a}, others
derived from QCD using effective field theory
\cite{epelbaum:00a,entem:03a,machleidt:11a}.

The most ambitious solutions of the three-particle scattering
problem only became possible by the emergence of larger and faster
computers in the late eighties of the last
century~\cite{gloeckle:96a,witala:88a,witala:88b,FRGL_3B}.
Calculation methods based on the solution of momentum space
integral equations could be achieved, both below and above
three-particle breakup, using real axis integration together with
spline interpolation and subtraction methods to handle the
singularities that exist along the momentum axis for real energy
 \cite{gloeckle:96a}.
The sole draw back that subsided for many years in momentum
space calculations was the ability to include the long range Coulomb
interaction between charged particles that was finally overcome in
2005 \cite{deltuva:05a,deltuva:05d}
using a very simple but efficient screening procedure of the
Coulomb interaction followed by the renormalization of the scattering
amplitudes, that leads to a fast convergence of the calculated
observables in terms of screening radius.

Unlike momentum space calculations, coordinate space methods based
on formally exact theories were able to include the Coulomb
interaction since the very early developments by using appropriate
boundary conditions, leading to the interpretation of the existing
data for charge particle reactions which are in general more
abundant, easy to measure and more accurate. This is the case for
proton-deuteron elastic scattering where numerically converged
solutions have been around for many years
 using Kohn variational principle together with  the hyperspherical harmonics
expansion method \cite{kievsky:01a} or Faddeev equations \cite{chen:89}.
 A number of benchmark calculations
\cite{FRGL_3B,Friar_nd_bench,kievsky:01b,deltuva:05b} have been
performed over the years and demonstrate that momentum space and
coordinate space methods provide equally accurate solutions for
neutron-deuteron (n-d) and proton-deuteron (p-d) elastic
scattering, both below and above three-particle breakup threshold.
Nevertheless, unlike momentum space calculations, the Achilles
heel of coordinate space calculations has been so far the
calculation of three-particle breakup observables due to numerical
difficulties associated with matching, at a chosen boundary where
the interactions are considered negligible, a complicated analytic
asymptotic wave function to the numerical solution of the
Schr\"{o}dinger equation calculated from inside out. Given the
oscillatory behavior of continuum wave functions for positive real
energy, this matching in six-dimension space is the source of
numerical inaccuracies and instabilities in the calculated
observables. Recently three different methods which allow to
solve the break-up problem without an explicit use of the asymptotic
form of system wave function, and thus apply bound-state-like
basis, have emerged with growing success.

The very first idea of using bound state solutions to solve
many-body scattering problems is already present in Wigners
R-matrix theory \cite{Wigner:46,Wigner:46_2,Wigner:47}. In this
approach the scattering observables were obtained from
configuration space solutions in the interaction region, which
were expanded in squared integrable basis functions and thus
without imposing the appropriate boundary conditions.\footnote{In
nuclear physics  the `phenomenological' R-matrix method is more
acknowledged as a technique to parametrize various
types of cross sections. However in other domains of physics
`calculable' R-matrix method is renown as a rigorous calculational
tool to derive scattering properties from the Schr\"{o}dinger
equation; see ref.~\cite{Descouvemont:10}  for a detailed
description.} {However, at least in the nuclear physics case, this
method was mostly used to solve scattering problems with only
binary (elastic and rearrangement) channels open. Furthermore this
method requires full} diagonalization of the Hamiltonian
representing matrix; thus, the resulting linear algebra problem
must be of limited size to remain treatable numerically. Therefore the
R-matrix method has been mostly
applied~\cite{Descouvemont:06,Descouvemont:10,Arai:11} to handle
simplified model problems (models with primitive interactions or
including approximate dynamics) rather than providing exact
solutions for few-body problems.


The first successful application of bound state methods to the
calculation of observables, both below and above breakup threshold
has been developed in Trento \cite{Efros:1994iq} to treat
processes that are driven by an external source such as a photon
or an electron. The photo disintegration of $\He$
\cite{LaPiana:2000jg,Golak:2002mw}, ${}^4\mathrm{He}$
\cite{Quaglioni:2004} and other light nuclei
\cite{Bacca:2004,Bacca:2004a,Efros:2007nq} or the electron
disintegration $(e,ep)$ of bound light clusters
\cite{Quaglioni:2005} have been calculated using the Lorentz
Integral Transform (LIT) which is the natural extension of an
original idea to calculate reaction cross sections with the help
of integral transforms. This method concentrates directly on
matrix elements instead of trying to calculate wave functions and
therefore avoids solving the Schr\"{o}dinger equation in the
continuum. Using clever integral transforms together with a chosen
Lorentz kernel, the LIT method can reduce a continuum problem to a
much less problematic bound-state-like problem. Nevertheless,
since the LIT method has not yet been used successfully to
calculate hadronic reactions, the solution of coordinate space
Faddeev-like equations \cite{faddeev:60a} above breakup has to
rely on other methods, {like} complex scaling where the
coordinates r are multiplied by a complex unit phase leading to
their rotation by an angle $\theta$. With an appropriate choice of
$\theta$ that depends on how the interaction behaves at large
distances one is able to convert the solution of a continuum
equation into a bound state like one. This method has been applied
to handle very diverse
problems~\cite{MKK01,SMK05,AMKI06,KKMTI10,KKWK11,KKMI13,LC11,La12},
can cope with realistic interactions~\cite{DFL12,LaFB13}, and
the results compare well with other calculation methods.


  If the three-particle scattering problem is complicated, the
  four-particle is even more so given the increased dimensionality
  \cite{yakubovsky:67,grassberger:67,fonseca:87}
(number of vector variables, partial waves and mesh size
  required for convergence). Therefore, for many years the
  four-nucleon scattering problem did not catch the attention of
  few-body physics due to both a combination of lack of available
  computer power to handle the dimensionality of the problem in its
  full glory, and also the ability to overcome the complicated
  singularity structure of the four-body Kernel in momentum space or
  the intricate boundary conditions in coordinate space for
  multichannel problems assymtotically. The lack of appropriate
  treatment of the Coulomb interaction was an additional drawback for
  momentum space calculations.

For the above mentioned reasons four-nucleon scattering results
with realistic force models such as AV18 emerged first through
coordinate space calculations but limited to single channel
problems assymtotically, such as $\nH$ and $\pHe$ reactions
\cite{viviani:01a,kievsky:08a,lazauskas:04a} where only the
elastic channel exists up to three-body breakup threshold, or
$\pH$ below the $\nHe$ threshold \cite{lazauskas:09a}. In that
region these reactions present a rich structure of resonances
\cite{tilley:92a} in different partial waves that have been well
identified in the literature and whose understanding in terms of
the underlying force models constitute a major unresolved
challenge for theory. More recent results show that adding a
three-nucleon three-body force such as Urbana IX \cite{urbana9} to
AV18 does not necessarily improve
\cite{lazauskas:04a,lazauskas:05a,fisher:06} the agreement with
the experimental data. As in the three-nucleon system, complex
scaling methods are now being used to calculate single channel
reactions above breakup threshold. Nevertheless the interactions
being used so far are still restricted to s-waves~\cite{LC11}.

Due to its inherent complexity, rich structure of resonances and
multitude of channels in both isospin $T=0$, $T=1$ and mixed
isospin the four-nucleon system constitutes an ideal theoretical
laboratory to test nucleon-nucleon (NN) force models. But for that
to be possible one needs to be able to solve numerically, over a
broad range of energy, the corresponding momentum or coordinate
space equations.

Given that the treatment of the Coulomb interaction between
protons became possible in momentum space calculations by using
the method of screening and renormalization mentioned above,
solutions of the Alt, Grassberger and Sandhas equations
\cite{grassberger:67,deltuva:07a} for the transition operators
have been done at energies below breakup threshold for a number of
realistic NN interactions such as AV18 \cite{wiringa:95a}, CD Bonn
\cite{machleidt:01a}, INOY04 \cite{doleschall:04a} and N3LO
\cite{entem:03a}. Because assymtotic boundary conditions are
naturally imposed by the way one handles the two-body
singularities, one could calculate cross sections and spin
observables for all two-body reactions ranging from $\nH$
\cite{deltuva:07a}, $\pHe$ \cite{deltuva:07b}, $\nHe$, $\pH$ and
$\dd$ \cite{deltuva:07c} elastic scattering to transfer reactions
such as ${}^3\mathrm{H}(p,n){}^3\mathrm{He}$,
${}^3\mathrm{H}(p,d){}^2\mathrm{H}$, and
${}^3\mathrm{He}(n,d){}^2\mathrm{H}$ \cite{deltuva:07c} and their
respective time reversal. In this energy range calculations were
done using real axis integration, spline interpolation, two-body
subtraction methods and Pad\'{e} \cite{baker:75a} summation of all
3N, 2N+2N and 4N amplitudes \cite{deltuva:07a}. This same approach
was used to study 3N- and 4N-force effects \cite{deltuva:08a} on
the above observables by using CD Bonn + $\Delta$
\cite{deltuva:03c} force model that extends the Hilbert space to
include NN-N$\Delta$ coupling in addition to NN-NN. Unlike
coordinate space methods, adding a static 3N-force to the
underlying 2N force constitutes a major stumbling block for
momentum space calculations that has not yet been resolved, except
for bound state calculations \cite{nogga:02a}.

Given the complex analytical structure of the four-body Kernel in
momentum space above breakup threshold, going beyond three-particle
threshold seemed for a while an impossible endeavor. Real axis subtraction methods did not work as long as the energy E was real. Using complex energy in the form of $Z = E + i\epsilon$, where $\epsilon$ is a finite quantity  \cite{uzu:03a}, was a mirage that only worked well when new weights for the integration mesh \cite{deltuva:12c} were developed that already take into account
the nature of the singularities. Great progress has recently been
achieved, leading for the first time to realistic state of the art
calculations of $\nH$ \cite{deltuva:12c} and
$\pHe$ \cite{deltuva:13c} elastic scattering up to 35 MeV
lab energy. Due to the complex energy method, integration on the real
momentum axis only faces quasi singularities that are accurately
calculated by the new weight scheme.

In conclusion, our collective experience tell us that the recent
developments we present in this manuscript may provide the
necessary tools to overcome serious difficulties in the solution
of the multiparticle scattering problem, both in coordinate and
momentum space calculations. Although the many-body scattering
problem in its full complexity may be yet decades away from an
exact numerical solution, a first step has already been taken by
combining LIT and Coupled-cluster methods to handle dipole
response of $^{16}O$ nucleus~\cite{Bacca:2013dma}. Furthermore,
either complex scaling, complex energy or continuum discretization
methods, we overview in this manuscript, require very limited
effort to be incorporated in conjunction with the most advanced
bound state techniques, like No-core shell
model~\cite{Barrett:2013ppnp} or Coupled-cluster
method~\cite{Hagen:2010prc}, enabling to handle many-body
collisions. We should mention recent very challenging developments
that combine the resonating group method and no-core shell
model~\cite{Navratil:2010}  to solve elastic scattering problems
beyond the $A=4$ case. However in the last approach dynamics of
the many-body system is still treated approximately. Due to the
limited scope of this review we chose to concentrate on fully
rigorous methods that enable solutions of the scattering problem
both below and above the three-particle breakup threshold.

In addition, the methods we review here may contribute to further
progress the solution of the three- and four-body problems
involving not just nucleons but also higher-body systems that
under given circumstances effectively exhibit three- or four-body
degrees of freedom. This is indeed the case in direct nuclear
reactions involving the scattering of a deuteron  or halo nucleus
from a nuclear target \cite{deltuva:07d,deltuva:09a} that can be
as light as a proton or as heavy as $^{208}\mathrm{Pb}$. Using
effective interactions such as optical potentials and
complementary structure information one may be able to make
predictions \cite{deltuva:09b,deltuva:09d} that may lead to the
extraction of important information from the data collected by
radioactive ion beam experiments. Progress in this area has been
slow over the years, although some advances have been made in the
last ten years. The major stumbling block in this endeavor is the
ability to reduce a many-body scattering problem to an effective
fewer-body one preserving unitarity and including the appropriate
structure effects that are needed to characterize the underlying
subsystems. Work in this direction was formulated years ago
\cite{polyzou:79a} but never successfully implemented.

In this paper we present six different techniques to solve the
multiparticle scattering problem using bound state techniques,
namely:  Lorentz Integral Transform  method (Section
\ref{sec:lit}), Techniques based on continuum-discretized states
(Section \ref{sec:cds}), Complex scaling method (Section
\ref{sec:csm}), Complex energy methods in configuration (Section
\ref{sec:cemc}) and momentum (Section \ref{sec:cemm}) spaces,
Momentum lattice technique (Section \ref{sec:mml}). We conclude
the paper by giving an outlook in Section
\ref{sec:out}.

%% file: Contrib_Jaume_cor.tex
\section{Lorentz integral transform \label{sec:lit}}

A genuine method  to  compute the scattering observables in terms
of bound state wavefunctions was proposed by V. Efros in
\cite{Efros_SJNP_1985}. There exist a recent very detailed review
on this method~\cite{Efros:2007nq};  therefore we restrict this section to a brief summary of the key ideas.

For  this purpose let us consider  the response function
$R(\omega)$ of a system driven by a Hamiltonian $H$  to some perturbation operator
$\hat{O}$ which is responsible for the energy transfer $\omega$
and for inducing transitions from its ground-state $\mid 0
\rangle$ with energy $E_{0}$, to an arbitrary states $E_n$ of its
spectrum \begin{equation}\label{Rom}
 R(\omega) = \sum_n \mid \langle n\mid    \hat{O} \mid 0 \rangle\mid^2
 \delta(E_n-E_0-\omega).
 \end{equation}
The sum appearing in the right hand side of this expression
involves all the states with an energy $E_n$ being coupled to the
ground state by the operator $\hat{O}$, and can thus include all
the continuum many body states.

The key point of this method  is to remark  that while computing the response function  by means of (\ref{Rom}) would require the {\it a priori} knowledge of
the full spectrum of $H$. Its integral transform with a Stieltjes kernel
\begin{equation}\label{SIT}
 \Phi_S(\sigma)=   \int_0^{\infty}   \frac{R(\omega)}  {\omega +
 \sigma}   \qquad \sigma>0, 
  \end{equation}
can be expressed as the expectation value of the inverse (shifted)
Hamiltonian on the perturbed vacuum, i.e.:
\begin{equation}\label{VEV_SIT}
 \Phi_S(\sigma) = \sum_n  {\mid \langle n\mid \hat{O} \mid 0 \rangle\mid^2 \over E_n-E_0 + \sigma}
 = \langle 0\mid  \hat{O}^{\dagger}  {1\over H-E_0+\sigma}   \hat{O}  \mid 0\rangle  =\langle S\mid {1\over H-E_0+\sigma}  \mid
 S\rangle,
 \end{equation}
with
\begin{equation}\label{S}
\mid S\rangle =\hat{O} \mid 0\rangle.
\end{equation}

This integral transform $\Phi_S(\sigma) $  can thus be easily obtained as  the scalar product
\begin{equation}\label{SP}
  \Phi_S(\sigma)    = \langle S\mid  \Psi_S(\sigma) \rangle,
 \end{equation}
where $\mid \Psi_S(\sigma) \rangle $ is a solution of the inhomogeneous Schr\"{o}dinger-like equation:
\begin{equation}\label{ISCE_S}
 (H-E_0+\sigma)  \mid \Psi_S(\sigma) \rangle   =    \mid S\rangle.
 \end{equation}

The initial bound  state $\mid0\rangle$
is transformed, by the action of  the perturbation  $\hat{O}$,  into a short range vector $\mid S \rangle$.
According to (\ref{SP}), the integral transform $ \Phi_S(\sigma)$  is an overlap between the final state, involving very complex many-body wave functions,
and this short range  vector $\mid S \rangle$. This overlap implies an effective  truncation of the asymptotic part of the final-state which makes it insensitive to its behaviour.
The integral transform is thus a quantity entirely determined by the inner region and keeps all the information of the asymptotics in the normalization constant
determined by the inhomogeneous term of eq. (\ref{ISCE_S}).

The interest of this approach lies in the fact that the solution
of (\ref{ISCE_S}) is unique, squared integrable, and can be
computed by using powerful bound state methods.
This is a
remarkable result in what allows to obtain a quantity  $R(\omega)$
which involves an infinity of  states, including many breakup
channels with highly non trivial boundary conditions, in terms of
easy to handle solutions $\mid \Psi_S(\sigma) \rangle$ with
trivial asymptotes. Even if these solutions must be known for a
continuous set of the parameter $\sigma$, the benefit is
substantial.

\bigskip
There is  an additional step before the physical observables
related to the quantity $R(\omega)$ can be calculated;  that is
the inversion of the integral equation (\ref{SIT}). This is
however a non trivial issue since it belongs to the so called "ill
conditioned problem".

This point, crucial for practical applications of the method, is addressed with some detail in \cite{Barnea:2009zu}.
It should be not understood as an anomaly in the mathematical sense but rather as a numerical
difficulty in inverting eq. (\ref{SIT}).
For an arbitrary, though invertible kernel, it can happen that very different
response functions would map into very close transform functions $\Psi_S(\sigma)$, thus making the inversion of (\ref{SIT}) more difficult.
This procedure, mathematically well defined, would be exact with an infinite precision
but would require a highly accurate calculation not free from instabilities.
The choice of an integral kernel is a compromise, to some extend empirical, to achieve
the inversion with a reasonable numerical cost as well as to ensure a practical independence of the results on the parameters $\sigma$.


It became soon clear that the inversion of
eq. (\ref{SIT}) with the Stieltjes kernel was not stable enough to
produce reliable results even for the simplest 2-body transitions,
like those related with the deuteron~\cite{Efros:1993xy}.

An essential improvement was done when replacing  the Stieljes kernel (\ref{SIT}) by the Lorentz one  \cite{Efros:1994iq}:
\begin{equation}\label{LIT}
  \Phi_L(\sigma_R,\sigma_I)=   \int_0^{\infty}   {R(\omega)  \over (\omega
  -\sigma_R)^2+\sigma_I^2}.
 \end{equation}
This form can be viewed as a representation of a $\delta$-function
that turns out to be suitable for most practical applications.

The equivalent of equation (\ref{VEV_SIT}) becomes now
\[  \Phi_L(\sigma_R,\sigma_I)  =   \sum_n  {\mid \langle n\mid  \hat{O} \mid 0\rangle\mid^2 \over (E_n-E_0 - \sigma_R)^2+\sigma_I^2}
= \langle 0\mid  \hat{O}^{\dagger}  {1\over
H-E_0-\sigma_R+i\sigma_I} \; {1\over H-E_0-\sigma_R-i\sigma_I}
\hat{O}  \mid 0\rangle,   \] that is
\[  \Phi_L(\sigma_R,\sigma_I)= \langle \Psi_L\mid  \Psi_L \rangle, \]
with
\begin{eqnarray*}
 (H-E_0-\sigma-i\sigma_I)  \mid \Psi_L \rangle   &=&    \mid
 S\rangle.
 \end{eqnarray*}
The use of the so called Lorentz Integral Transform  (LIT) ensured a good control in the inversion
process \cite{Barnea:2009zu} and has been at the origin
 of substantial developments during the last
 twenty years with applications to a great number of  perturbation-induced reactions in nuclear physical.

\bigskip

The numerical set up was first worked out   by  computing the
total deuteron longitudinal response function \cite{Efros:1994iq}
and has since then  been successfully extended to compute
inclusive inelastic reactions of nuclei with $A=3, 4, 6, 7$
induced by electroweak process (photons, electrons and neutrinos).
The LIT method has  also been extended to compute exclusive cross
section in reactions with more than two particles in the final
state \cite{LaPiana:2000jg}, like for instance
\begin{itemize}
\item Longitudinal cross section in
$d(e,e'N)N$~\cite{LaPiana:2000jg}
 \item Ab initio calculation of
the $^4$He(e,e'd)d reaction~\cite{Leidemann:1994} \item Two-body
photodisintegration of $^4$He involving many rearrangement and
break-up channels~\cite{Quaglioni:2004,Quaglioni2007370}
\begin{eqnarray*}
 \gamma + ^4He  &\to& p + ^3H \cr
                               &\to& n + ^3He \cr
                                &\to& d+ d \cr
                                &\to& p+n +d \cr
                                &\to& p+p+n+n
\end{eqnarray*}
\item $^4He(e, e'p)^3H$ reaction with full final-state
interactions~\cite{Quaglioni:2005}
\end{itemize}

Finally, the LIT method can be applied in conjunction with the
other few-body approaches to reach until now unexplored regions. A
recent work merged the LIT  to coupled-cluster
method~\cite{Hagen:2010prc} and obtained a first principles
computation of the giant dipole resonance in $^{16}$O
\cite{Bacca:2013dma}.

\begin{figure}[h,t]
\begin{center}
\epsfig{file=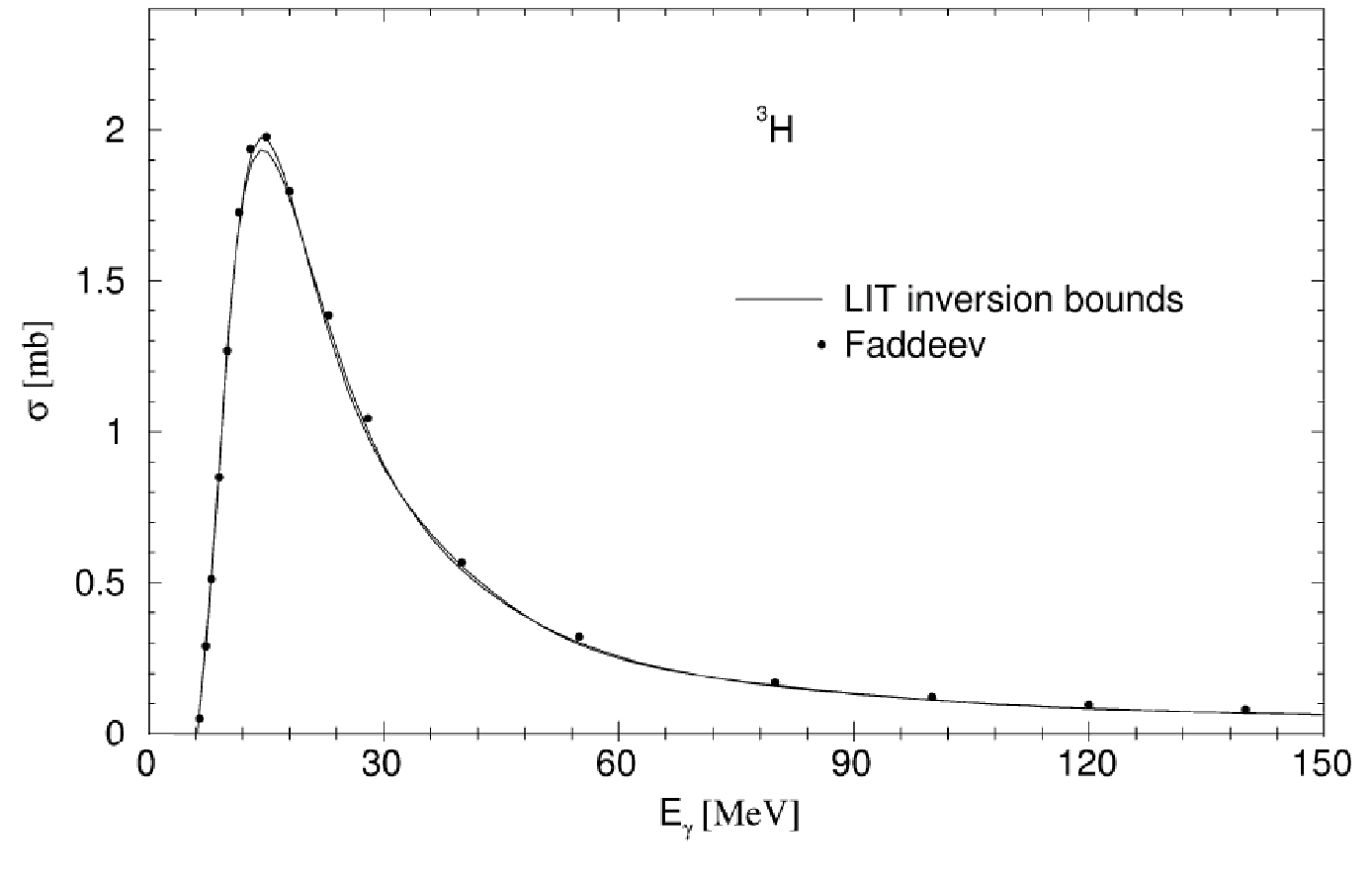,scale=0.5}
\caption{ Comparison between Faddeev results (dots) and Lorentz
Integral transform (solid lines) in the  three-nucleon
photodisintegration total cross section with modern realistic two-
and three-nucleon forces \cite{Golak:2002mw}. The  two LIT curves
correspond to the small uncertainties in the inversion of the
transform (\ref{LIT}).} \label{fig1}
\end{center}
\end{figure}

When the comparison is possible, the LIT results have been found
in perfect agreement with direct few-body calculations, like
the solution of Faddeev-Yakubovsky equations incorporating the
full complexity of the continuum states. An example is provided by
the three-nucleon photodisintegration cross section displayed in
Fig. \ref{fig1}. The results taken from ref. \cite{Golak:2002mw} were obtained employing the Argonne AV18~\cite{wiringa:95a} nucleon-nucleon potential
and the Urbana UIX~\cite{urbana9}
three-nucleon force. The LIT case includes also the Coulomb
force. Faddeev results are indicated by dots and LIT ones by two
curves corresponding to the uncertainties in the inversion
 of eq.(\ref{LIT}). One can see perfect agreement between the
two approaches, whereas the uncertainties of LIT method are very
small.

The Lorentz Integral Transform  represents nowadays the most
efficient approach to challenge  many-body systems in the
continuum. It allows to calculate perturbation induced reaction
observables without an explicit use of many-body continuum wave
functions. As pointed out in the Introduction, the calculation of
these wave functions is made difficult by the existence of many
open channels in the continuum, in particular those involving
many-body break up reactions. In configuration space calculations the corresponding
solutions face the problem of implementing
the appropriate boundary conditions. In momentum space, these
difficulties are translated into a very complex structure of
singularities. In practice exact solutions,  in the framework
of Faddeev-Yakubovski or AGS equations,  are presently  limited to
A=3 and A=4 systems. The LIT approach therefore constitutes an
efficient way to circumvent the boundary condition problem and go
well beyond A=4 systems. The interested reader will find a
comprehensive review of this approach and  a complete reference
list in \cite{Efros:2007nq}.

Although the usual formulation of the LIT method is based on
perturbation theory, it  can be in principle extended to non
perturbative reactions. This possibility was already present in
its initial formulation (Sec. 4 of ref.~\cite{Efros_SJNP_1985})
and has been also emphasized in Sec. 2.3 of a more recent review
\cite{Efros:2007nq}. Nevertheless, until now, the existing results
are however limited to electroweak processes. It would be
interesting to demonstrate its applicability to reactions driven by the strong interaction alone, like the simplest elastic $n+^3H$
scattering, or neutron induced break-up reactions.

\newpage

\section{Scattering amplitude calculation using continuum-discretized states \label{sec:cds}}

As mentioned in the Introduction, the very first idea for using
squared integrable basis function to solve the scattering problem
was already present in the Wigner's R-matrix approach, back in the
forties \cite{Wigner:46,Wigner:46_2,Wigner:47, Lane:1948zh}. In
the nuclear physics community  this approach was extensively  used
to parameterize scattering experiments rather than to compute
effectively ab initio scattering observables
\cite{Descouvemont:10}.

The same ideas have been revised  from a slightly different
perspective, some years later, in a series of papers by Harris
starting from ref. \cite{Harris_PRL19_1967}. In the simple case of a one
channel Schr\"{o}dinger equation it can be formulated as follows.

We consider the stationary scattering solution of the
Schr\"{o}dinger equation at energy E
\begin{equation}\label{SCH}
H \Psi = E \Psi,
\end{equation}
in the form
\begin{equation}\label{Sol}
  \Psi = c_1 \Psi_1(E) + c_2 \Psi_2(E) + \Phi,
\end{equation}
where $\Psi_i$ are two independent  asymptotic solutions at energy
E of the free equation and $\Phi$ is an unknown function to be
determined, as well as the coefficients $c_i(E)$. The scattering
observables are obtained from the ratio of these coefficients, in
a way depending on the particular choice of the asymptotic
solutions $\Psi_i$. If, for a three-dimensional solution of the
Schr\"{o}dinger equation, one choses for instance
\[ \Psi_1 = e^{ i\vec{k}_0\cdot\vec{x} }  \qquad  \Psi_2=  \frac{e^{i\mid\vec{k}\mid r }}
{r},
\]
the ratio would give the scattering amplitude
\begin{equation}\label{Samp}
f  (  \vec{k}_0 \cdot \vec{k} )   = {c_2\over c_1}.
\end{equation}

The problem remains to determine in an efficient way this ratio
and this was the main result in ref. \cite{Harris_PRL19_1967}. By
inserting (\ref{Sol}) into  (\ref{SCH}) one finds
\begin{equation}\label{SCH2}
c_1(E-H)\Psi_1(E) + c_2(E-H)\Psi_2(E) + (E-H)\Phi =0.
\end{equation}
It is worth noticing that, in the asymptotic region, $\Phi$
vanishes only if the coefficients $c_i$ are those giving the right
asymptotic behavior of the scattering solution. However  each of
the three terms in equation (\ref{SCH2}) vanishes in the
asymptotic  region for arbitrary values of $c_i$ . It is then
natural to project this equation on a basis  of squared integral
functions.

Let us denote as $B=\{b_i\}_{i=1,\ldots,d}$ a finite basis set,
not necessarily orthogonal, and let us diagonalize the Hamiltonian
$H$ on this d-dimensional basis by solving the generalized
eigenvalue equation
\[ (H - \epsilon N) \phi =0, \]
with the d-dimensional matrix
\begin{eqnarray*}
H_{ij} &=& <b_j\mid H \mid b_i> \cr N_{ij} &=& <b_j \mid b_i>.
\end{eqnarray*}
The solutions thus obtained ($\epsilon_n,\phi_n$) --  assumed by
simplicity non degenerate -- provides an approximate spectral
representation of the Hamiltonian
\[ H\approx H'= \sum_{n=1}^{d} \mid \phi_n>\epsilon_n <\phi_n\mid,
\]
which can be done  more and more accurate  by increasing the
number of states in the basis set.

Let us now project the equation (\ref{SCH2}) onto one of these
eigenvectors $\phi_m$
\begin{equation}\label{SCH3}
c_1<\phi_m\mid E-H\mid\Psi_1(E)> + c_2<\phi_m\mid E-H\mid
\Psi_2(E)> + <\phi_m\mid E-H\mid \Phi> =0.
\end{equation}
The key point of the method is to remark that by choosing E to be
the corresponding eigenvalue $\epsilon_m$
 the term depending on the unknown function $\Phi$ vanishes and one gets
 \begin{equation}\label{SCH3}
c_1 <\phi_m\mid \epsilon_m-H\mid\Psi_1(\epsilon_m)> + c_2
<\phi_m\mid \epsilon_m-H\mid \Psi_2(\epsilon_m)>  =0.
\end{equation}

The ratio of the coefficients $c_i$ obtained in this way provides
an accurate estimation of the scattering amplitude
eq.(\ref{Samp}) at the energy $E=\epsilon_m$, i.e. it is obtained
as a ratio of the integral expressions:
 \begin{equation}\label{SCH4}
f  (  \vec{k}_0 \cdot \vec{k} )   = {c_2\over c_1} =
-\frac{<\phi_m\mid \epsilon_m-H\mid\Psi_1(\epsilon_m)>}{
<\phi_m\mid \epsilon_m-H\mid \Psi_2(\epsilon_m)>}.
\end{equation}

By interpolating scattering between the ones corresponding to
calculated eigenvalues or by adjusting the basis set one can
access to an accurate description of the scattering process at any
desired energy E. In ref. \cite{Harris_PRL19_1967} the method was
successfully applied to compute the phase shifts of a two-body
problem interacting via a Yukawa potential and in a subsequent
work to the scattering of electrons on atomic  hydrogen
\cite{Michels_Harris_PRL19_1967}. The method  can be in principle
generalized to a many-particle system.

\bigskip
Such bound state approach to scattering solutions gained
an increasing interest in recent years. They are all based in
using the continuum discretize states to obtain the phase shifts
at the corresponding energies. In ref. \cite{Suzuki_NPA_09} a method
was developed that makes use of the Green's function formalism  to
obtain the integral representation of the phase shifts  in terms
of a bound state basis set. It was  applied to calculate p-n and
n-$^4$He scattering observables using semirealistic interactions.

An extension of this approach to coupled channel problems  is
presented in ref. \cite{Suzuki_NPA_10}. The method is based on the
introduction of a confining potential in the external region which
enables to find a number of independent energy-degenerate
solutions corresponding to the number of coupled-channels.


A generalization of the Harris et al. approach and a step further
in the complexity of the calculation was  achieved in ref.
\cite{Pisa_PRC_12}. These authors obtain the scattering amplitude
as a ratio of integral expressions like in eq.(\ref{SCH4}),  but
they improve this result by taking benefit from the Kohn
variational principle
\cite{Barletta_PRL103_2009,Kievsky_PRC81_2010}. Using a basis of
bound-state-like wave functions  the scattering matrix
corresponding to the n-d (A=3) and p-$^3$He (A=4) scattering was
computed for  realistic Hamiltonians \cite{Pisa_PRC_12}. This required the
extension of the Kohn variational principle to the coupled-channel
case. The construction of the energy degenerate bound-state-like
wave functions belonging to the continuum spectrum of the
Hamiltonian is discussed.

In summary, the method of continuum-discretized states, with
several extensions to long range forces and coupled-channel
problems, provided already accurate results in systems up to A=4.
It has reached now a maturity to be extended in the near future to
study more complex systems. Implementation of this method should
be straightforward in conjunction with any bound state technique
to handle single channel collisions. On the other hand validity of
the method should yet be proved for the energies above the
three-particle break-up threshold.

%% file: Contrib_Rimas.tex
\section{Complex Scaling Methods in configuration space \label{sec:csm}}

The solution of the scattering problem in configuration space is a
very difficult task both from  formal (theoretical) as well as
 computational points of view. The principal difficulties
arise from the complex asymptotic behavior of the system wave
function.

Configuration space wave functions in the asymptotic region combine
as many outgoing waves as there are open channels. Moreover, for
the break-up channels, which include more than two charged
particles, outgoing wave solutions are not even known
analytically. The resulting asymptotic form for three-charged
particles has been elucidated in ref.~\cite{AM93}, however its
complexity has discouraged all the efforts to use
it explicitly as a boundary condition for solving the
Schr\"{o¨}dinger equation. Methods that enable the solution of the
scattering problem without using the asymptotic form
of the wave function  present enormous benefits.

The complex scaling~(CS) technique has been introduced already
during the World War II by D.R. Hartree et al.~\cite{HMN46,Co83}
in the study of the radio wave propagation in the atmosphere. D.R.
Hartree et al. were solving second order differential equations
for complex eigenvalues. In practice, this problem is
equivalent to the one encountered in the search for resonance positions in quantum
two-particle collisions. In the late sixties Nuttal and
Cohen~\cite{NC69} proposed a very similar technique to treat the
generic scattering problem for short range potentials. Few years
later Nuttal even employed this method to solve the three-nucleon
scattering problem above breakup threshold~\cite{MDN72}.
Nevertheless these pioneering works of Nuttal have been mostly
forgotten, while based on Nutall's work and the mathematical
foundation of Baslev and Combes~\cite{BC71} the original method of
Hartree has been recovered in order to calculate resonance
eigenvalues in atomic physics~\cite{Ho83,Mo98}. Such an omission
is mostly due to the fact that short range potentials may gain a
highly untrivial structure after the complex scaling
transformation~\cite{REB81,Witala:3n,LC05_3n} is applied, while
for the Coulomb potential this transformation is trivial.

Only recently a variant of the complex scaling method  based on
the spectral function formalism has been presented by Kat\={o},
Giraud et al.~\cite{MKK01,GK03,GKO04} and applied in the works of
Kat\={o} et al.~\cite{MKK01,SMK05,AMKI06,KKMTI10,KKWK11,KKMI13}.
This variant will be described in detail in the end of this
section. On the contrary in the later works of Kruppa et
al.~\cite{KSK07} as well as in the works of two of us (J.C. and
R.L.)~\cite{LC11,La12} the original idea of Nuttal and Cohen is
elaborated.

\subsection{Two-body problem}

\subsubsection{Short range, exponentially bound, interactions}
The idea of Nuttal and Cohen~\cite{NC69} can briefly be formulated
as follows. The Schr\"{o}dinger equation is recast into its
inhomogeneous (driven) form by splitting the wave function into
the sum $\Psi (r)=$ $\Psi ^{sc}(r)+\Psi^{in}(r)$, where
the incident (free) $\Psi ^{in}(\mathbf{r}%
)=\exp (i\mathbf{k}\cdot \mathbf{r})$ wave is separated.
The remaining untrivial part of the system wave function  $\Psi ^{sc}(\mathbf{r%
})$ describes the scattered waves and may be found by solving a
second-order differential equation with an inhomogeneous term:
\begin{equation}
\lbrack E-\hat{H_{0}}-V(\mathbf{r})]\Psi ^{sc}(\mathbf{r})=V(\mathbf{r}%
)\Psi ^{in}(\mathbf{r}).  \label{Schro_2B}
\end{equation}
The scattered wave in the asymptote is represented by an outgoing
wave $\Psi^{sc}\sim \exp (ikr)/r,$ where $k=\sqrt{2\mu E}/\hbar$
is the wave number for the relative motion. If one scales all the
particle coordinates by a constant complex factor, i.e. $r^\theta
_{i}=e^{i\theta }r_{i}$ with $Im(e^{i\theta })>0$, the
corresponding scattered wave $\overline{\Psi}^{ sc}(\mathbf{r})$
will vanish exponentially $\sim \exp (-kr\sin \theta )$ as
particle separation $r$ increases. Moreover if the interaction is
of short range -- exponentially bound with the longest range
$\eta^{-1}$ -- then after complex scaling the right hand side
of eq.~(\ref{Schro_2B}) also tends to zero at large $r$, if :
\begin{equation}
\tan \theta <\eta /k.  \label{Cond_shrp}
\end{equation}%
From here we introduce the notation $f^\theta (r)=f(re^{i\theta })$
for the complex-scaled functions. The complex scaled driven
Schr\"{o}dinger equation becomes:
\begin{equation}
\lbrack
E-e^{-i2\theta}\hat{H_{0}}-V^\theta(\mathbf{r})]\overline{\Psi}^{sc}
(\mathbf{r})=V^\theta(\mathbf{r})(\Psi^{in})^\theta(\mathbf{r}).
\label{Schro_2B_cs}
\end{equation}%
If the condition in eq.~(\ref{Cond_shrp}) is satisfied, the former
inhomogeneous equation may be solved by using a compact basis to expand $\overline{%
\Psi }^{sc}(\mathbf{r})$, thus by employing standard bound-state
techniques.

 There are two ways to extract scattering information
from the obtained solutions $\overline{\Psi }^{sc}(\mathbf{r})$.
One method is based on the
asymptotic behavior of the outgoing waves, where the
scattering amplitude $f_{k}(\hat{r})$
is extracted in a similar way as the asymptotic
normalization coefficient from the bound-state wave function, that is,
 by matching asymptotic behavior of the solution:
\begin{equation}
\overline{\Psi }^{sc}(\mathbf{r})=f_{k}(\widehat{k})e^{-i\theta
}\exp (ikre^{i\theta })/r.  \label{local_2B_0}
\end{equation}

The other well known alternative is to use an integral relation which
one gets after applying the Green's
theorem~\cite{KSK07,LC11,HK12}:
\begin{eqnarray}
f_{k}(\widehat{k}) &=&-\frac{2\mu }{\hbar ^{2}}e^{i3\theta }\int
(\Psi^{in \ast})^\theta
(\mathbf{r})V^\theta(\mathbf{r})\left[\overline{\Psi }%
^{sc}(\mathbf{r})+(\Psi^{in})^\theta
(\mathbf{r})\right]d^{3}r \\
&=&-\frac{2\mu }{\hbar ^{2}}e^{i3\theta }\int (\Psi^{in
\ast})^\theta
(\mathbf{r}){V^\theta}(\mathbf{r})\overline{\Psi }%
^{sc}(\mathbf{r})d^{3}r-\frac{2\mu }{\hbar ^{2}}\int (\Psi ^{in} (\mathbf{r}))^{\ast}V(%
\mathbf{r}){\Psi }^{in}(\mathbf{r})d^{3}r . \label{integral_2B}
\end{eqnarray}%
In the second relation one has separated the Born term which may
be evaluated without performing complex scaling. The $(\Psi^{in
\ast})^\theta(\mathbf{r})$ term is obtained by applying the
complex-scaling operation on the complex-conjugate function
$(\Psi^{in }(\mathbf{r}))^\ast$.

\subsubsection{Presence of the long range interaction}

Let us consider the case where the interaction  has an additional
long-range term
$V(\mathbf{r})=V_{s}(\mathbf{r})+V_{l}(\mathbf{r})$, where
$V_{s}(\mathbf{r})$ is exponentially bound and $V_{l}(\mathbf{r})$
is long-ranged. The CS method can be generalized to treat this
problem if for the long range term $V_{l}(\mathbf{r})$ the
incoming wave solution $\Psi _{l}^{in}(\mathbf{r})$ is analytic
and can be extended into the complex
r-plane~\cite{KSK07,Elander_CSM,LC11}.  For the
Coulomb case $V_l(\mathbf{r})=\frac{\hbar^{2}\eta_C }{\mu r}$ the
incoming wave solution is well known and is usually expanded in
terms of the regular Coulomb functions $F_{\ell}(\eta_C,kr)$. Then
one is left to solve the equivalent driven Schr\"{o}dinger
equation:
\begin{equation}
\lbrack E-e^{-i2\theta }\hat{H_{0}}-V^\theta(\mathbf{r})]\overline{%
\Psi }_{s}^{sc}(\mathbf{r})=V^\theta_{s}(\mathbf{r})(\Psi_{l}^{in})^\theta(%
\mathbf{r}).  \label{Schro_2B_lr}
\end{equation}%
The inhomogeneous term on the right hand side of the former
equation is moderated by the short-range interaction term;
therefore it is exponentially bound  if the condition (\ref{Cond_shrp}) is fulfilled by the short range potential
$V_{s}(\mathbf{r})$.

One may establish a relation equivalent to the
eq.(\ref{integral_2B}) in order to determine the
long-range-modified short-range interaction amplitude
$f_{k,s}(\widehat{k})$ :
\begin{eqnarray} \nonumber
f_{k,s}(\widehat{k}) &=&-\frac{2\mu }{\hbar ^{2}}e^{i3\theta }\int
(\Psi^{in \ast}_l)^\theta(\mathbf{r})V^\theta_{s}(\mathbf{r})%
\overline{\Psi}_{s}^{sc}(\mathbf{r})d^{3}r \\
&&-\frac{2\mu }{\hbar ^{2}}\int (\Psi _{l}^{in}(\mathbf{r}))^{\ast }V_{s}(%
\mathbf{r}){\Psi }_{l}^{in}(\mathbf{r})d^{3}r.
\end{eqnarray}%
The total scattering amplitude $f_{k}(\widehat{k})$ is a sum of
the short-range one and the scattering amplitude due to the
long-range term  alone $f_{k,l}(\widehat{k})$ :
\begin{equation}
f_{k}(\widehat{k})=f_{k,s}(\widehat{k})+f_{k,l}(\widehat{k}).
\end{equation}


\subsection{N-body problem}

\subsubsection{Short range interactions\label{sec:cs_nbody_sr}}

Here we present the general formalism to treat the collisions of
two multiparticle clusters. Lets consider two clusters $a$ and $b$ formed by
$N_{a}$ and $N_{b}$ particles (with $N_{a}+N_{b}=N$) whose
binding energies are $E_{a}$ and $E_{b}$ respectively. The
relative kinetic energy of the two clusters in the center of mass
frame is $E_{a,b}=E_{c.m.}-E_{a}-E_{b}=$ $\hbar
^{2}k_{a,b}^{2}/2\mu _{a,b}$. Then the incoming wave takes the
following form:
\begin{equation}
\Psi _{a,b}^{in}(\mathbf{k}_{a,b},\mathbf{r}_{i,a}\mathbf{,r}_{j,b}\mathbf{,r%
}_{a,b})=\psi _{a}(\mathbf{r}_{i,a})\psi _{b}(\mathbf{r}_{j,b})\exp (i%
\mathbf{k}_{a,b}\cdot \mathbf{r}_{a,b}),
\end{equation}
where $\psi _{a}(\mathbf{r}_{i,a})$ and $\psi _{b}(\mathbf{r}_{j,b})$
represent
bound state wave functions of the clusters $a$ and $b$ respectively, with $\mathbf{%
r}_{i,a}(\mathbf{r}_{j,b})$ defining internal coordinates of the clusters, while $%
\mathbf{r}_{a,b}$ is a vector connecting the centers of mass of the
two clusters.

As previously, one writes the Schr\"{o}dinger equation in its
inhomogeneous form and applies the complex scaling on all the
coordinates, getting:
\begin{equation}
\lbrack
E-e^{-i2\theta}\hat{H}_0-\sum\limits_{m<n}{V^\theta_{mn}}\left(
\mathbf{r}_{m}\mathbf{-r}_{\mathbf{n}}\right)]\overline{\Psi }%
_{a,b}^{sc}(\mathbf{r}_{i,a},\mathbf{r}_{j,b}\mathbf{%
,r}_{a,b})=\left[\sum\limits_{i\in a;j\in b}V^\theta_{ij}\left(
\mathbf{r}_{i}\mathbf{-r}_{\mathbf{j}}\right)\right]
(\Psi_{a,b}^{in})^\theta(\mathbf{r}_{i,a},\mathbf{r}_{j,b},\mathbf{%
r}_{a,b}). \label{Schr_dr_nbody}
\end{equation}
The term $\overline{\Psi}_{a,b}^{sc}(\mathbf{r}_{i,a},\mathbf{r}_{j,b},\mathbf{%
r}_{a,b})$ contains only complex-scaled outgoing waves in the
asymptote and thus is formally bound exponentially. Therefore, as
long as the right hand side of the last equation is bound, it might
be solved using square a integrable basis set to express the
scattered part of the wave function
$\overline{\Psi}^{sc}_{a,b}(\mathbf{r}_{i,a}\mathbf{,r}_{j,b}\mathbf{,r}_{a,b})$.

\begin{figure}[tb]
\par
\begin{center}
\begin{minipage}[t]{8 cm}
\epsfig{file=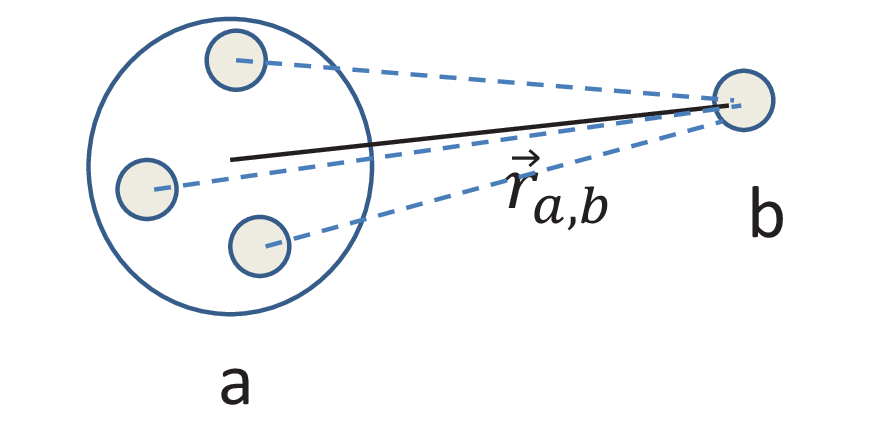,scale=0.5}
\end{minipage}
\begin{minipage}[t]{16.5 cm}
\caption{Directions of the interaction terms between the two
multiparticle clusters (a and b) do not coincide exactly with the
wave vector $\vec{r}_{a,b}$ connecting their centers of
mass.\label{fig_cl_ab}}
\end{minipage}
\end{center}
\end{figure}

However the inhomogeneous term of eq.~(\ref{Schr_dr_nbody}) is not
necessarily exponentially bound even if all the interaction terms
are bound. This is due to the fact that, unlike for 2-body case,
the directions of the interaction terms do not coincide with the
wave vector connecting the center of mass of the two clusters as shown in
Fig.~\ref{fig_cl_ab}. Still one may demonstrate that the
inhomogeneous term remains bound if an additional condition is
fulfilled~\cite{LC11}:
\begin{equation}
\label{cond_lim_nb}
\tan \theta <\min (\sqrt{\frac{B_{i\in a}}{E_{a,b}}\frac{m_{i}(M_{a}+M_{b})}{%
(M_{a}-m_{i})M_{b}}},\sqrt{\frac{B_{j\in b}}{E_{a,b}}\frac{m_{j}(M_{a}+M_{b})%
}{(M_{b}-m_{j})M_{a}}}),
\end{equation}
where $B_{i\in a}$ is the i-th particle removal energy from the
cluster $a$ and $M_{a}$ is a total mass of the cluster $a$. The
last condition implies additional limit on the complex scaling
angle $\theta $ to be used. For a system of equal mass particles
this limit does not have much effect and becomes important only
well above the break-up threshold $\left\vert E_{a,b}\right\vert
>>B_{i\in a}$ (or $\left\vert E_{a,b}\right\vert
>>B_{j\in b}$ respectively). Even at high energies
this limit is not so constraining,  since the exponent of the
scattered wave becomes proportional to $\sqrt{E_{a,b}}$ and
therefore one may achieve the same speed of convergence by
employing smaller complex scaling angle $\theta$ values. On
the other hand the condition in eq.~(\ref{cond_lim_nb})  may
become strongly restrictive for  the mass-imbalanced systems when
one considers light-heavy-heavy components.

The scattering observables are easy to calculate using the Green's
theorem. The elastic scattering amplitude is obtained from
\begin{eqnarray} \nonumber
f_{a,b}(\hat{k}_{a,b})
&=&-\frac{2\mu}{\hbar^{2}}e^{i3(N-1)\theta}\int({\Psi }_{a,b}^{in
\ast})^\theta \left[\sum_{i\in a;j\in b} V^\theta_{ij} \left(
\mathbf{r}_{i}\mathbf{-r}_{\mathbf{j}}\right)
\right] \overline{\Psi }_{a,b}^{sc}\prod_{k=1}^{N-1}d^{3}x_{k} \\
&&-\frac{2\mu }{\hbar ^{2}}\int (\Psi _{a,b}^{in})^{\ast }\left[
\sum\limits_{i\in a;j\in b}V_{ij}\left( \mathbf{r}_{i}\mathbf{-r}_{\mathbf{j%
}}\right) \right] \Psi
_{a,b}^{in}\prod\limits_{k=1}^{N-1}d^{3}x_{k}, \label{eq:el_ampl}
\end{eqnarray}
where $x_{k}$ represents the internal Jacobi coordinates of the system.
Thus one integrates over the full N particle volume by excluding the
center of mass one.

 Inelastic amplitudes are provided by
\begin{eqnarray} \nonumber
f_{a,b->c,d}(\widehat{k}_{a,b},\widehat{k}_{c,d}) &=&-\frac{2\mu }{\hbar ^{2}%
}e^{i3(N-1)\theta}\int ({\Psi }_{c,d}^{in ^{\ast }})^\theta\left[
\sum\limits_{i\in c;j\in d}V^\theta_{ij}\left( \mathbf{r}_{i}\mathbf{-r}_{\mathbf{%
j}}\right) \right] \overline{\Psi }_{a,b}^{sc}\prod%
\limits_{k=1}^{N-1}d^{3}x_{k} \\
&&-\frac{2\mu }{\hbar ^{2}}\int (\Psi _{c,d}^{in})^{\ast }\left[
\sum\limits_{i\in c;j\in d}V_{ij}\left( \mathbf{r}_{i}\mathbf{-r}_{\mathbf{%
j}}\right)\right] \Psi
_{a,b}^{in}\prod\limits_{k=1}^{N-1}d^{3}x_{k}. \label{eq:inel_ampl}
\end{eqnarray}

Finally, based on expressions provided
in refs.~\cite{Glockle_book,gloeckle:96a}, the break-up amplitude into
three clusters $(c,d,f)$ is given by:
\begin{eqnarray} \nonumber
f_{a,b->c,d,f}(\widehat{k}_{a,b},K_{cdf},\widehat{k}_{c,d},\widehat{k}%
_{cd,f}) &=&-\frac{2\mu }{\hbar ^{2}}e^{i3(N-1)\theta }\int ({%
\Psi }_{c,d,f}^{in \ast})^\theta {V_{cdf}^\theta}\overline{\Psi }%
_{a,b}^{sc}\prod\limits_{k=1}^{N-1}d^{3}x_{k} \\
&&-\frac{2\mu }{\hbar ^{2}}\int (\Psi _{c,d,f}^{in})^{\ast }
V_{cdf}\Psi _{a,b}^{in}\prod\limits_{k=1}^{N-1}d^{3}x_{k}.
\label{eq:brc_ampl}
\end{eqnarray}
In the last equation the interaction term is
\begin{equation}
V_{cdf}=\sum_{i<j}V_{ij}-\sum_{(m<k)\in c}V_{mk}-\sum_{(m<k)\in
d}V_{mk}-\sum_{(m<k)\in f}V_{mk}.
\end{equation}

\subsubsection{Presence of the long range interaction}

One may try to include the long-range interaction in a similar
manner as it has been done for the 2-body case. To this aim one
should separate the incoming wave $\left( \Psi
_{a,b}^{in}\right)_{l}$ modified by the residual long-range
interaction term $V_{l}(\mathbf{r}_{a,b})$. In particular, for the
charged projectile-target system it is natural to subtract the
residual Coulomb interaction between the colliding clusters $a$
and $b$ with $V_{l}(\mathbf{r}_{a,b})=Z_aZ_b/\mathbf{r}_{a,b}$,
where $Z_a$ and $Z_b$ are respective charges of the projectile and
target respectively.
\begin{equation}
\lbrack E-e^{-i2\theta
}\hat{H}_{0}-\sum\limits_{m<n}{V_{mn}^\theta}\left(
\mathbf{r}_{m}\mathbf{-r}_{\mathbf{n}}\right)]\left({%
\Psi }_{a,b}^{sc}\right) _{l}(\mathbf{r}_{i,a}\mathbf{,r}%
_{j,b}\mathbf{,r}_{a,b})=\left[ \sum\limits_{i\in a;j\in
b}{V_{ij}^\theta}\left(
\mathbf{r}_{i}\mathbf{-r}_{\mathbf{j}}\right)
-{V_{l}^\theta}(\mathbf{r}_{a,b})\right] {\left(\Psi
_{a,b}^{in}\right)^\theta_{l}}(\mathbf{r}_{i,a}\mathbf{,r}%
_{j,b}\mathbf{,r}_{a,b}).
\end{equation}
However the residual interaction term $\left[ \sum\limits_{i\in
a;j\in b}{V_{ij}^\theta}\left(
\mathbf{r}_{i}\mathbf{-r}_{\mathbf{j}}\right)
-{V_{l}^\theta}(\mathbf{r}_{a,b})\right]$, appearing on the right
hand side of this equation, still retains some higher-order terms
which may not converge exponentially. In particular, for the
aforementioned case of the charged clusters (Coulomb case) this
interaction term retains higher order Coulomb multipolar terms
starting with $1/r_{a,b}^{3}$, which are due to the possible
polarization of the projectile-target. Therefore the right-hand
side of the last equation is not exponentially bound but contains
long-range slowly diverging terms\footnote{Strictly speaking,
similar diverging terms appear in the expressions equivalent to
eqs.({\ref{eq:el_ampl}-\ref{eq:brc_ampl}}) for the scattering
amplitudes.}. Nevertheless these multipolar terms are weak, even
compared to the subtracted long-range interaction term (as
discussed for $Z_aZ_b/\mathbf{r}_{a,b}$ Coulomb case) and should
represent only mild corrections in the far asymptote region.
Therefore if a system is dominated by the strong short-range
interaction terms one might eventually consider screening these
multipolar terms due to the residual long-range interaction. Such
a procedure is used in obtaining the results presented in the next
section for three-body systems interacting via Coulomb plus
short-range nuclear potentials finding no consequences on the
final result.

\subsubsection{External probes}

There is a group of problems in physics where the system is
initially in a bound state and gets subsequently excited to the continuum by an external source that is considered as a
perturbation. In particular, it concerns reactions led by
electro-magnetic and weak probes.
 In this case one is interested in evaluating the strength or
response function given in lowest order perturbation theory as
\begin{equation}
S(E)=\sum_{\nu }\left\vert \left\langle \Psi _{\nu }\left\vert
\widehat{O}\right\vert \Psi _{0}\right\rangle \right\vert
^{2}\delta (E_{\nu }-E_{0}-E),
\end{equation}
where $\hat{O}$ is the perturbation operator which induces the
transition from
the bound-state $\Psi _{0},$ with ground-state energy $E_{0}$, to the state $%
\Psi _{\nu }$ with  energy $E_{\nu }$. Both wave functions are
solutions of the same Hamiltonian $H$ . The energy is measured
from some standard value, e.g., a particle-decay threshold energy.
When the excited state is in the continuum, the label $\nu $ is
continuous and the sum must be replaced by an integration.
Furthermore the final state wave function $\Psi _{\nu }$ may have
complicate asymptotic behavior in configuration space if it
represents continuum states. On the other hand the expression may \ be
rewritten by avoiding summation over the final states
\begin{eqnarray}
S(E) &=&\left\langle \Psi _{0}\left\vert \widehat{O}^{\dag }\delta
(H-E_{\nu
})\widehat{O}\right\vert \Psi _{0}\right\rangle  \\
&=&-\frac{1}{\pi }Im\left\langle \Psi _{0}\left\vert
\widehat{O}^{\dag }G(E_{\nu }+i\varepsilon )\widehat{O}\right\vert
\Psi _{0}\right\rangle =-\frac{1}{\pi }Im\left\langle \Psi
_{0}\left\vert \widehat{O}^{\dag }\right\vert \Phi _{\nu
}\right\rangle \label{Strenght_func}
\end{eqnarray}
with
\begin{equation}
(H-E_{\nu })\Phi _{\nu }=\widehat{O}\Psi _{0}.
\end{equation}
The right hand side of the former equation is compact, damped by the bound-state $%
\Psi _{0}$ wave function. The wave function $\Phi _{\nu }$
asymptotically contains only outgoing waves. Therefore the
last inhomogeneous equation may be readily solved using complex
scaling techniques
\begin{equation}
(H^{\theta }-E_{\nu })\overline{\Phi} _{\nu }=\widehat{O}^{\theta
}\Psi _{0}^{\theta }.
\end{equation}
One must just use the complex scaled expressions for the right
hand side of the equation. The complex-scaled bound state wave
function $\Psi_{0}^{\theta }$ is obtained by solving the bound state
problem using the complex-scaled Hamiltonian
\begin{equation}
(H^{\theta }-E_0)\Psi _{0}^{\theta }=0,
\end{equation}
and finally
\begin{eqnarray}
S(E) =-\frac{1}{\pi }Im\left\langle \Psi _{0}^{\theta }\left\vert
(\widehat{O}^{\dag })^\theta\right\vert \overline{\Phi} _{\nu
}\right\rangle.
\end{eqnarray}

\subsubsection{Complex scaled Green's function method  \label{CS_GF}}
\begin{figure}[tb]
\par
\begin{center}
\begin{minipage}[t]{8 cm}
\epsfig{file=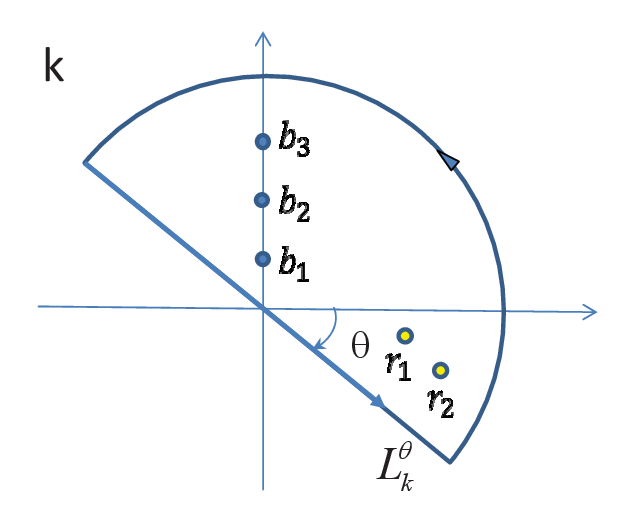,scale=0.5}
\end{minipage}
\begin{minipage}[t]{16.5 cm}
\caption{The Cauchy integral contour in the momentum plane for the
completeness relation of the complex scaled Hamiltonian. The
$b_1,b_2,..$ and $r_1,r_2,..$ represent the bound and resonant
poles respectively. \label{fig_cauchy}}
\end{minipage}
\end{center}
\end{figure}
A slightly different procedure to obtain the physical solution of
the complex scaled Hamiltonian has been proposed by Kat\={o},
Giraud et al.~\cite{MKK01,GK03,GKO04}. In the last paper the
completeness relation of the Berggren~\cite{Be73} has been proved
for the complex scaled Hamiltonian solutions representing bound,
resonant as well as single- and coupled-channel scattering states.
This completeness relation can be formulated for the Cauchy
integral contour in the momentum plane as demonstrated in
fig.~\ref{fig_cauchy}, as:
\begin{equation}
\mathbf{1}=\sum_{B}\left\vert \chi _{B}^{\theta }\right) \left(
\chi _{B}^{\theta }\right\vert +\sum_{R}^{n_{R}^{\theta
}}\left\vert \chi_{R}^{\theta }\right) \left( \chi _{R}^{\theta
}\right\vert +\int_{L_{k}^{\theta }}dk_\theta\left\vert \chi
_{k_{\theta }}\right) \left( \chi _{k_{\theta }}\right\vert,
\label{eq:GF_CSM}
\end{equation}
here $\chi _{B}^{\theta }$ and $\chi _{R}^{\theta }$ are the
complex scaled bound and resonant state wave-functions
respectively. Only the resonant states encircled by the semicircle
rotated by angle $\theta $ must be considered. The remaining continuum
states $\chi _{k}^{\theta }$ are located on the rotated momentum
axis $L_k^\theta$ (see Fig.~\ref{fig_cauchy}). One should mention
that the definition of the complex scaled bra- and ket-states for
the non-Hermitian $H^\theta$ are different from the usual ones of
the Hermitian Hamiltonian. For $H^\theta$ one expresses the
bra-state as the bi-conjugate solution of the ket-state. In
practice, for the discrete (resonant and bound) states we can use
the same wave functions for the bra- and ket-states; for the
continuum states the wave function of the bra-state is given by
that of the ket-state divided by the S-matrix. This is the reason
we use different notation to designate the complex scaled bra- and
ket-states $\left\vert \chi^{\theta }\right)$, instead of the commonly
accepted notation $\left\vert \chi\right>$ for solutions of Hermitian Hamiltonians.

Using the former completeness relation, the complex scaled Green's
function is written
\begin{equation}
\mathcal{G^\theta}
(E,\mathbf{r},\mathbf{r'})=\sum_{B}\frac{\left\vert
\chi_{B}^{\theta }(\mathbf{r})\right) \left( \chi _{B}^{\theta
}(\mathbf{r'})\right\vert}{E-E_B} +\sum_{R}^{n_{R}^{\theta
}}\frac{\left\vert \chi _{R}^{\theta }(\mathbf{r})\right) \left(
\chi _{R}^{\theta }(\mathbf{r'})\right\vert}{E-E_R}
+\int_{L_{k}^{\theta }}dk_\theta\frac{\left\vert \chi _{k_{\theta
}}(\mathbf{r})\right) \left( \chi _{k_{\theta
}}(\mathbf{r'})\right\vert}{E-E_\theta}, \label{eq:Greens_func}
\end{equation}
where $E_B$ and $E_R=(E_r-\frac{i}{2}\Gamma)$ are the energy
eigenvalues of the bound and relevant resonant states
respectively. Variables $\mathbf{r}$  reflect all the internal
coordinates of the multiparticle system under consideration. By
plugging the last relation into the eq.~(\ref{Strenght_func}), one
finally gets:
\begin{eqnarray}
S(E) &=&S_B(E)+S^\theta_R(E)+S^\theta_k(E), \\
S_B(E)&=&-\frac{1}{\pi }Im\sum_{B}\frac{\left( \Psi_{0}^{\theta
}\right\vert(\widehat{O}^\dag)^\theta\left\vert \chi_{B}^{\theta
}\right) \left( \chi _{B}^{\theta }\right\vert \widehat{O}^\theta
\left\vert \Psi_{0}^{\theta }\right)}{E-E_B},\\
\label{Strenght_func_gf}
 S^\theta_R(E)&=&-\frac{1}{\pi
}Im\sum_{R}^{n_{R}^\theta}\frac{\left( \Psi_{0}^{\theta
}\right\vert(\widehat{O}^\dag)^\theta\left\vert \chi_{R}^{\theta
}\right) \left( \chi _{R}^{\theta }\right\vert \widehat{O}^\theta
\left\vert \Psi_{0}^{\theta }\right)}{E-E_R},\\
S^\theta_k(E)&=&-\frac{1}{\pi }Im\int_{L^\theta_k}\frac{\left(
\Psi_{0}^{\theta }\right\vert(\widehat{O}^\dag)^\theta\left\vert
\chi_{k_\theta }\right) \left( \chi _{k_\theta}\right\vert
\widehat{O}^\theta \left\vert \Psi_{0}^{\theta
}\right)}{E-E_\theta}.
\end{eqnarray}
In practice (numerical solution) one works with a finite basis;
then the last term containing the integration is replaced by the sum
running over all the complex eigenvalues representing continuum
pseudo states. All the eigenvalues are obtained as solutions of
the complex scaled Hamiltonian with a pure outgoing wave boundary
condition (i.e. exponentially converging ones due to complex
scaling).

The obtained total strength function $S(E)$ should be independent
of the angle $\theta$ employed in the calculation. Furthermore the
strength function component $S_B(E)$, as well as its partial
components due to separate bound states are also independent of
$\theta$. The partial components of $S^\theta_R(E)$,
corresponding the same narrow resonance, also turn out to be
independent of $\theta$ as long as the angle $\theta$ is large
enough to encircle this resonance. However if the resonance is
large enough and is not encircled by the contour $L^\theta_k$, its
contribution to the strength function is reabsorbed by the
pseudo-continuum states in the $S^\theta_k(E)$ term. This feature
has been clearly demonstrated in ref.~\cite{AMKI06} for a
chosen 2-body example.

Relation~(\ref{Strenght_func_gf}) offers an unique feature to
separate the contributions of the resonant and bound states in the
strength function. The contributions of the narrow resonances
should not depend on the angle $\theta$, if the angle $\theta$ is
large enough to encircle a considered resonance.

\begin{figure}[!]
\begin{center}
\mbox{\epsfxsize=7.2cm\epsffile{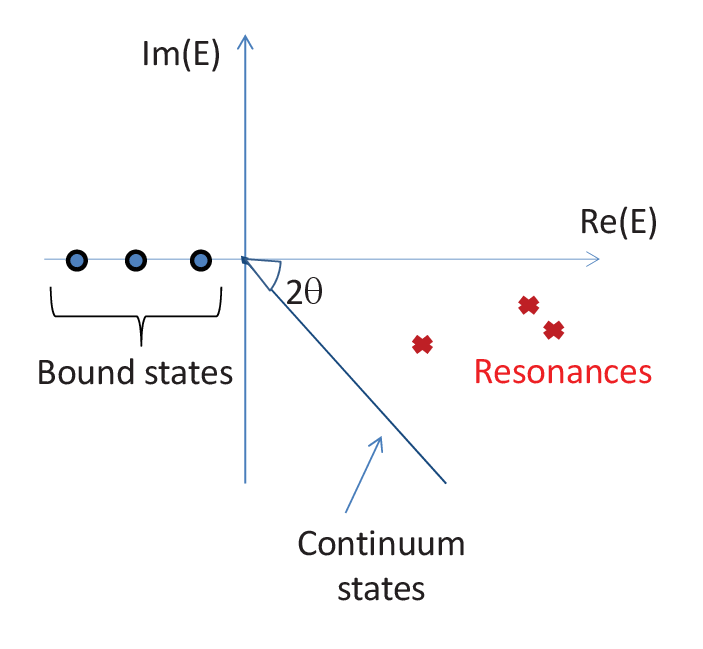}}
\mbox{\epsfxsize=10.2cm\epsffile{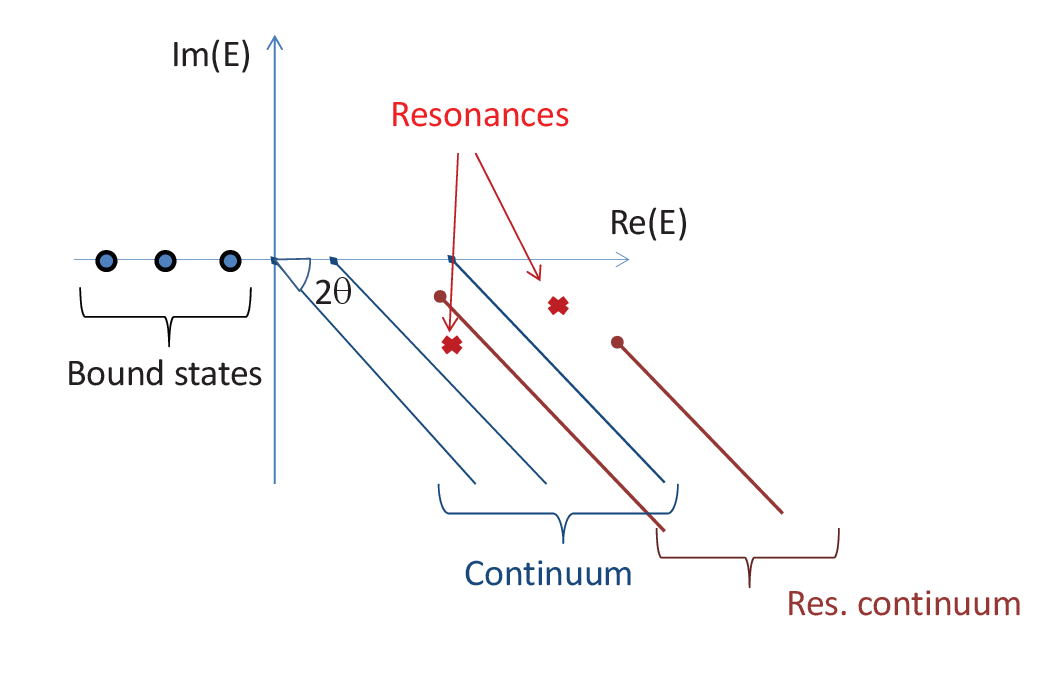}}
\end{center}
\caption{(Color online) Schematic representation of the
eigenvalues of the complex scaled Hamiltonian, $H^\theta$ ,
according to the theorem of Balslev and Combes~\cite{BC71}. For a
two-body system (left-pane) bound states are obtained as negative
real energy eigenvalues, continuum-pseudostates are rotated by
angle $2\theta$, resonances inside $2\theta$ branch may also be
obtained. For a many-body system (right pane) several rotated
continuum branches exist associated with bound and resonant
subcluster thresholds.} \label{fig:res_dep}
\end{figure}

For the sake of simplicity, the contour depicted in the
Fig.~\ref{fig_cauchy} represents the simplest 2-body case. Still
all of the relations presented remain valid for the many-body system;
 one only should keep in mind that the obtained spectra may have a
much more complicated structure. Following the Balslev and Combes
theorem~\cite{BC71} the eigenvalues of the complex-scaled two-body
Hamiltonian, which are associated with the bounded wave function,
splits into three categories:  bound state eigenvalues situated on
the negative horizontal energy axis, pseudo-continuum states
scattered along the positive energy axis rotated by angle
$2\theta$ and some eigenvalues representing the resonances whose
eigen energies satisfy the relation -$arg(E)<2\theta$ (see left
plane of Fig.~\ref{fig:res_dep}). For the many-body system bound
states will be situated on the horizontal part of the energy axis,
situated below the lowest system separation into multiparticle
cluster threshold (see Fig.~\ref{fig:res_dep}).
Pseudo-continuum states will scatter along the $2\theta$-lines
projected from each possible separation threshold. In addition one
will have $2\theta$-lines projected from the "resonant
thresholds", where one or more sub-cluster is resonant. Finally,
many-body resonance eigenvalues will represent discrete points
inside the semicircle making angle $2\theta$ with the real energy axis
and derived from the lowest threshold.

\subsection{Scattering amplitude via Greens-function method \label{SA_GF}}

The first application of the CS Green's function method to
calculate scattering phaseshifts has been realized by using the
continuum level density (CLD) formalism. One starts with the CLD
definition as
\begin{equation}
\Delta (E)=-\frac{1}{\pi} \textrm{Im}\left(\textrm{Tr}[
G(E)-G_0(E]\right), \label{eq:cld_0}
\end{equation}
with $G(E)=(E-H)^{-1}$ and $G_0(E)=(E-H_0)^{-1}$ being full and free
Green's functions, respectively. In principle, the former expression
may be generalized to the scattering of two complex clusters.
Then $H_0$, besides the kinetic energy, should include
interactions inside separate clusters, whereas $H$ includes all
the interaction terms in the two-cluster system. Thus CLD expresses the
effect from the interactions connecting the two clusters. When the
eigenvalues of $H$ and $H_0$  are obtained approximately
($\epsilon_i$  and $\epsilon^0_i$ respectively) within the
framework of including a finite number of the basis functions (N), the
discrete CLD is defined:
\begin{equation}
\Delta(E)_N=\sum_i\delta(E-\epsilon_i)-\sum_j\delta(E-\epsilon^0_j).
\label{eq:cld_1}
\end{equation}
The CLD is related to the scattering phaseshift
\begin{equation}
\Delta(E)=\frac{1}{\pi}\frac{d\varphi(E)}{dE}, \label{eq:cld_2}
\end{equation}
and thus one can inversely calculate the phaseshift~($\varphi$) by
integrating the last equation obtained as a function of energy.
These equations are difficult to apply for real Hamiltonians, as
one will necessarily confront the singularities present in
eq.~(\ref{eq:cld_0}-\ref{eq:cld_1}). However by using CS
expressions for the Green's functions, these singularities are
avoided and replaced by smooth Lorentzian functions. By
plugging in CS Green's function expression~(\ref{eq:Greens_func})
into~(\ref{eq:cld_1}) and after some simple algebra one gets:
\begin{equation}
\Delta(E)_N=\overline{\rho}_N^\theta(E)-\rho_N^{(0)\theta}(E)
 \label{cld_3}
\end{equation}
and
\begin{equation}
\overline{\rho}_N^\theta(E)=\frac{1}{\pi}\sum_R^{n_R^\theta}\frac{1}{(E-E_R)}+\frac{1}{\pi}\sum_k^{N-n_R^\theta-n_B}\frac{1}{(E-E^\theta_k)}.
\label{eq:cld_rhop}
\end{equation}
In the last expression $E_R$  and $E^\theta_k$ are the eigenvalues
of the full CS Hamiltonian $H^\theta$ representing resonant and
continuum states respectively. One should pay attention that the
sum over bound-states in the last expression is dropped. The term
$\rho_N^{(0)\theta}$ is equivalent to
$\overline{\rho}_N^\theta(E)$ obtained for the CS free
Hamiltonian $H_0^\theta$; this term contains only pseudo-continuum
states aligned along $2\theta$-lines pointing out from the
scattering thresholds (see Fig.~\ref{fig:res_dep}).

There exists however a much more straightforward way to calculate
scattering observables using CS Green's function~\cite{KSK07}.
Indeed, as one may see in eqs.~(\ref{eq:el_ampl}-\ref{eq:brc_ampl}),
the scattering amplitude naturally splits into the Born (trivial) and
the remaining (untrivial) term. The Born term may be calculated only
knowing the bound state wave functions of the incident and outgoing
subsystems. The nontrivial part of the scattering amplitude
requires knowledge of the CS scattered wave function $
\overline{\Psi }_{a,b}^{sc}$ at a given scattering energy $E$. This
part of the system wave-function is easily expressed using the CS
Green's function of eq.~(\ref{eq:Greens_func}):
\begin{equation}
\overline{\Psi }_{a,b}^{sc}(r)=\int d^3r' \mathcal{G^\theta}
(E,\mathbf{r},\mathbf{r'}) ({\Psi }_{a,b}^{in})^\theta(r'),
\label{eq:green_wf}
\end{equation}
where r and r' are all the internal coordinates in the
n-body system.

As demonstrated in ref.~\cite{KSK07}, the two approaches, the one
presented in subsection~\ref{sec:cs_nbody_sr} based on the
solution of differential equations with an inhomogenious term and
the one based on CS Green's function via
equations~(\ref{eq:cld_2}-\ref{eq:cld_rhop}), are fully
equivalent. I.e., if one employs the same numerical technique to
solve the differential equations with an inhomogenious term of the type
shown in eq.~(\ref{Schr_dr_nbody}) or one calculates eigenvalues
of the respective $H^\theta$ to approximate the  CS Green's
function in eq.(\ref{eq:Greens_func}), one will find identical
values for the scattering amplitude. On the contrary, the CLD
procedure to extract the scattering phaseshifts (or the amplitude
eventually) is not fully equivalent. Based on our limited
experience in the 2-body sector we found that the scattering
phaseshifts calculated using
expressions~(\ref{eq:Greens_func}) and (\ref{eq:green_wf}), are more
accurate than those obtained through
expressions (\ref{eq:cld_2}) through (\ref{eq:cld_rhop}) based on CLD formalism.

It should be noted that a full spectral decomposition of
$H^\theta$ is required to express CS Green's function in
eq.~(\ref{eq:Greens_func}) and to evaluate the scattering
amplitudes. The scattering amplitude, except in the case of
resonant scattering, is not determined by one or a few dominant
eigenvalues.\footnote{One should notice however, that if one tries to
approximate the phaseshifts using only the few eigenvalues that are
closest to the scattering energy, then the CLD formalism provides
better convergence than the
relations~(\ref{eq:cld_2}-\ref{eq:cld_rhop}).} This may turn out
to be a crucial obstacle in applying CS Green's function method in
studying many-body systems, since the resulting algebraic
eigenvalue problem becomes too large to be fully diagonalised. In
this case the original prescription of Nuttall, described in the
subsection~\ref{sec:cs_nbody_sr}, turns out to be strongly
advantageous. The last prescription resides on a single solution
of a linear-algebra problem, allowing one to employ the iterative
methods (without explicit storage of the matrix elements) to solve
a resulting large-scale problem.

On the other hand the CS Green's function formalism provides a clear
physical interpretation of the  scattering observables in terms of bound, resonant and continuum states. Furthermore, the same CS
Green's function expression is used to describe both collision
processes as well as system response to different perturbations
(like systems response to EM or weak field), thus providing a
solid ground to study correlations between different physical
observables.

\bigskip
Finally, one should discuss some technical aspects of the CS
method, which may hamper its successful implementation. First CS
implies complex arithmetics and non-Hermitian matrices already for
the problems involving only binary scattering channels. However
some linear-algebra methods used in numerical calculations are
limited to real Hermitian matrices. In the CS method one works with
the analytical potentials extended to the complex r-plane.
However, as pointed out in refs.~\cite{REB81,Witala:3n,LC05_3n} not all
the potentials behave well under complex scaling. In particular,
short-range potentials become oscillatory and even start diverging
for large $\theta$ values. Therefore, in numerical calculations it is advisable to keep the angle $\theta$ values small to
guarantee the smoothness of the potential after complex scaling in order to allow the numerical treatability of the
problem~\cite{Witala:3n,LC05_3n}. On the other hand the far
asymptote of the complex-scaled outgoing wave solution is
proportional to $exp(-k_xr_x\sin{\theta})$, where $k_x$ is a wave
vector corresponding to the last open-channel (channel with the
lowest free energy for the reaction products). Thus large angle
$\theta$ values are required to damp efficiently outgoing wave
solution if calculations are performed close to the threshold
(small $k_x$ value). This last fact makes it difficult to use CS in
exploring energy regions close to open thresholds.



\bigskip

\bigskip

\subsection{Results}

McDonald and Nuttall were the first  to apply the complex scaling
method to treat the scattering problem in $A>2$  system, already
back in 1972. They have studied neutron-deuteron scattering at
 neutron laboratory energies up to 24 MeV. Nucleon-nucleon
interaction in spin singlet and triplet channels has been
described by a single Yukawa-term potential, whose parameters have
been adjusted to reproduce the low-energy two-nucleon observables:
deuteron binding energy, singlet and triplet scattering length as
well as singlet effective range. Regardless of the simplicity of the
employed interaction, McDonald and Nuttall have managed to point
out the great difference in doublet and quartet inelastic
parameters, in particular demonstrating that the deuteron resists
 to breakup in the quartet channel. Furthermore
strong sensitivity of the doublet channel to the nature of the
NN-interaction has been revealed. The calculated neutron-deuteron
scattering length in the doublet channel, however, turned out to be too
large, as a result of the strongly overbound
triton\footnote{Roughly at the same time it has been observed in
numerical calculations by Phillips~\cite{Ph68} the existence of an
almost linear correlation between the triton binding energy and
neutron-deuteron scattering length. Now this correlation is renown
as the Phillips-line.}. This effect is undoubtedly due to the softness
of the employed NN interaction. In spite of these rather
encouraging results, the developments of McDonald and Nuttall have stopped.

Only in the late nineties has the complex scaling method been revisited
for scattering calculations while trying to apply it to Coulombic
systems~\cite{MCRB97,MCR97,BRILMC01}. Still, due to the dominance
of the long-range interaction, the direct approach described above does
not hold and thus a variant based on the exterior complex scaling
has been developed~\cite{Si79}. One should mention however that it
is extremely difficult to apply the exterior complex scaling method
to non-central or non-local interactions~\cite{Si79} as encountered
in nuclear physics.
\begin{figure}[!]
\begin{center}
\mbox{\epsfxsize=9.cm\epsffile{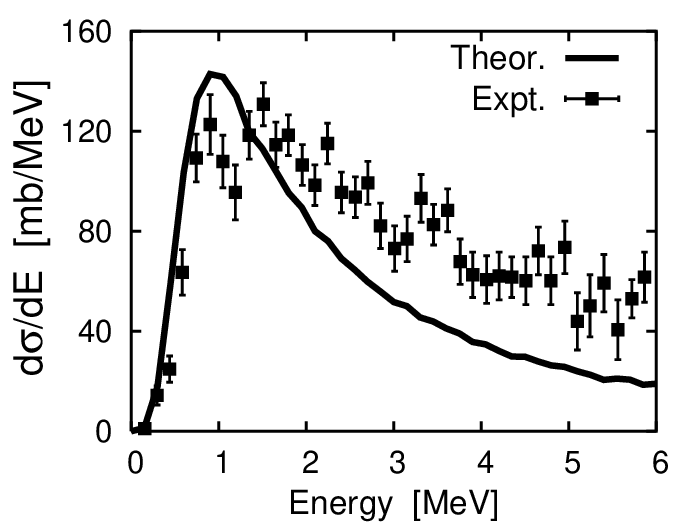}}
\mbox{\epsfxsize=9.cm\epsffile{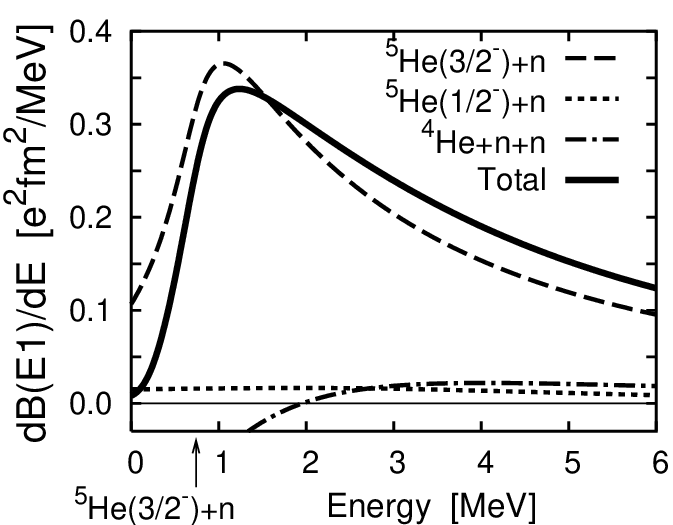}}
\end{center}
\caption{Strength distribution of the E1 disintegration of $^6He$
nucleus. In the left pane model results~\cite{AMKI06} are compared
with the experimental data of~\cite{AAAB99} using a secondary
$^6He$ ion beam of 240 MeV/nucleon incident on carbon and lead
targets. The calculated cross section is convoluted with respect
to experimental data. In the right pane the total strength (full
line) is split into contribution of two-body and three-body
channels. } \label{fig:He6}
\end{figure}

Interest in the CS method vis-$\grave {\rm {a}}$-vis nuclear reactions has
been revived by the work of Kat\={o} et
al.~\cite{MKK01,GK03,GKO04}. The method based on the spectral
decomposition described in section \ref{CS_GF} has been applied
by Kat\={o} et al.~\cite{SMK05,AMKI06} mostly in analyzing the EM
response of  two-neutron halo nuclei. In particular, E1 and E2
Coulomb breakup of $^6$He and  $^{11}$Li nuclei has been
studied, using a semi-microscopical three-body model. In such a
model $^6$He or  $^{11}$Li nuclei are represented by two neutrons
attached to the $^4$He or $^9$Li cores respectively. The core
cluster ($^4$He or $^9$Li) is considered to be in its ground
state,  and only the interactions between the two halo
neutrons and halo neutron-core are explicitly considered. The
three-body wave function is antisymmetric with respect to the last
two neutrons. The Pauli principle between the core and halo
neutrons is mimicked using the orthogonality condition model through the pseudo-potential method of Kukulin et al.~\cite{KKVS86}.
In Fig. \ref{fig:He6} the strength distribution of the E1
transition for $^6$He, as obtained in ref.~\cite{AMKI06}, is presented.
In the last calculation
 the microscopic KKNN potential~\cite{KKNN79}   and the
effective Minnesota potential~\cite{TLT78} have been used to
represent $V_{^4He-n}$  and $V_{n-n}$ interactions, respectively.
Such a simplistic model allows a rather accurate description of the
experimental data. Furthermore, from the obtained results one may
conclude the dominance of the sequential $^6\textrm{He}\rightarrow
^5\textrm{He +n} \rightarrow ^4\textrm{He +n +n}$ process in the
Coulomb breakup of $^6\textrm{He}$. This reaction proceeds mostly
through $J=3/2^-$ and $J=1/2^-$ resonances of $^5\textrm{He}$.
This demonstrates the importance of the CS Green's function method,
which provides a clear physical interpretation of the scattering
observables in terms of bound, resonant and continuum states.

In a later work by the same group of Japanese
scientists~\cite{KKMI13}  the EM breakup of $^{11}\textrm{Li}$ has been studied in an extended three-body model, by representing $^9\textrm{Li}$
core as a coupled cluster including $2p-2h$ excitations.

In ref.~\cite{KKWK11} the aforementioned three-body model has also been
 used to study deuteron elastic scattering on $^4\textrm{He}$
as well as $^2\textrm{H}(^4\textrm{He}, \gamma)^6\textrm{Li}$
radiative capture reaction.

On the contrary, in the studies by two of us (R.L. and J.C.)~\cite{LC11,La12},
the original complex scaling method of Nuttall and Cohen,
described in section ~\ref{sec:cs_nbody_sr},  is elaborated. The
few-body problem is solved for the complex scaled Faddeev-Yakubovski
equations. In particular the validity of the CS method has been
demonstrated for n+$^2\textrm{H}$ scattering above the deuteron
breakup threshold, by comparing results with the ones obtained
using direct configuration and momentum space methods~\cite{LC11}.
Furthermore the validity of the CS method has been demonstrated for
systems which interact via optical potentials that have an
absorbing-imaginary part~\cite{DFL12,LaFB13}. As a test case, the
n+p+$^{12}$C system has been considered within a three-body model.
In this system the n-p interaction was described using the realistic
AV18 model~\cite{wiringa:95a}. The interaction between the neutron
(proton) and the $^{12}$C core was simulated by the optical
potential~\cite{CH89}. Elastic p$+^{13}$C, d+$^{12}$C  as well as
inelastic p$+^{13}$C$\rightarrow$d$+^{12}$C cross sections have
been calculated for 30 MeV deuteron (or 30.6 MeV proton) laboratory
energy, which is above n+p+$^{12}$C breakup threshold. Excellent
agreement with a direct momentum space calculations based on AGS
equations and the Coulomb-screening method~\cite{deltuva:05d}, has
been obtained.

\begin{figure}[!]
\begin{center}
\mbox{\epsfxsize=8.5cm\epsffile{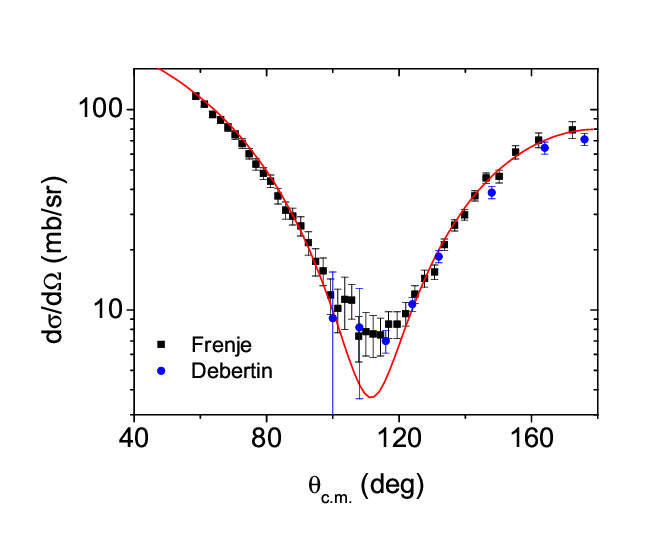}}
\mbox{\epsfxsize=8.5cm\epsffile{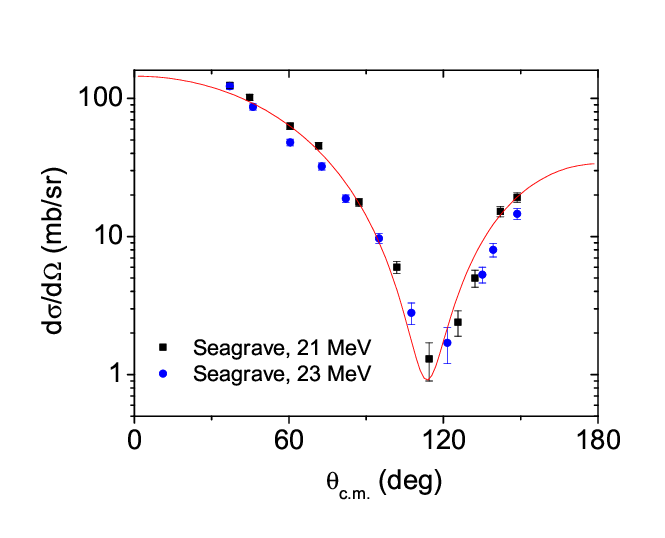}}
\end{center}
\caption{Calculated n+$^3$H elastic differential cross-sections
for neutrons of lab. energy 14.4 MeV (left pane) and 22.1 MeV
(right pane) compared with the experimental results of Frenje et
al.~\cite{Frenje_nt}, Debertin et al.~\cite{Debertin} and Seagrave
et al.~\cite{Seagrave}.} \label{fig:nt_MT}
\end{figure}

Lately the CS method has been applied to solve the four-nucleon
scattering problem~\cite{La12} for total isospin $T=0$ and $T=1$
channels. S-wave spin-dependent MT I-III potential was employed to
mimic the nucleon-nucleon interaction but ignoring the Coulomb repulsion
between the protons. The four-nucleon system has been studied both
above 3-body (the N+N+(NN) case) and 4-body (N+N+N+N) breakup
thresholds. Results have been compared with the ones obtained
using momentum space complex-energy method giving excellent
agreement. Reasonable agreement with the experimental data
of refs.~\cite{Frenje_nt,Debertin,Seagrave} has been found for
n+$^3$H scattering above the 4-body breakup threshold as shown in
Fig.~\ref{fig:nt_MT}. Even better description of the
experimental data can be found if realistic interactions are
used, as pointed out in ref.~\cite{deltuva:12c}.
 The isospin $T=0$ channel has been found to be very sensitive to the nucleon-nucleon interaction input and thus requires a more
realistic
 model in order to reproduce the experimental data.

\newpage
\section{Complex energy method in configuration space \label{sec:cemc}}

Effective range theory is one of the most popular tools in
analyzing and describing scattering processes. This fact clearly
indicates that  analytical continuation methods can be
successfully applied to circumvent the well-known difficulties in
solving scattering problems. Already in the mid-sixties Schlessinger
and Schwartz~\cite{SS66} proposed a continuation method to
calculate the elastic scattering amplitude from the results obtained
in the negative energy region. The idea of Schlessinger has been
generalized to complex energy  by McDonald and
Nuttall~\cite{MDN69}.  The starting point for this method is the
inhomogeneous Scr\"{o}dinger equation:
\begin{equation}
\lbrack E_c-\hat{H}_0-\sum\limits_{m<n}{V_{mn}}\left(
\mathbf{r}_{m}\mathbf{-r}_{\mathbf{n}}\right)]\overline{\Psi }%
_{a,b}^{sc}(E_c,\mathbf{r}_{i,a},\mathbf{r}_{j,b}\mathbf{%
,r}_{a,b})=\left[\sum\limits_{i\in a;j\in b}V^\theta_{ij}\left(
\mathbf{r}_{i}\mathbf{-r}_{\mathbf{j}}\right)\right]
\Psi_{a,b}^{in}(E'_c,\mathbf{r}_{i,a},\mathbf{r}_{j,b},\mathbf{%
r}_{a,b}). \label{Schr_dr_nbodyt}
\end{equation}
One solves this equation for a complex energy $E_c=|E_c| e^{2i
\theta}$ with a positive imaginary part (i.e. $\theta<\pi/2$). The
last condition on the angle $\theta$ allows to avoid the cut
along the real-energy axis, while the outgoing wave solutions $\overline{\Psi }%
_{a,b}^{sc}(E_c,\mathbf{r}_{i,a},\mathbf{r}_{j,b}\mathbf{%
,r}_{a,b})$ fall exponentially at large distances. One may use
integral relations formulated in subsection~\ref{sec:cs_nbody_sr}
in order to evaluate scattering amplitudes for complex energies in
the upper half-plane. Physical amplitudes, corresponding the real
energy values, might be extrapolated from the amplitude values
obtained for the complex energies.
\begin{figure}[tb]
\par
\begin{center}
\epsfig{file=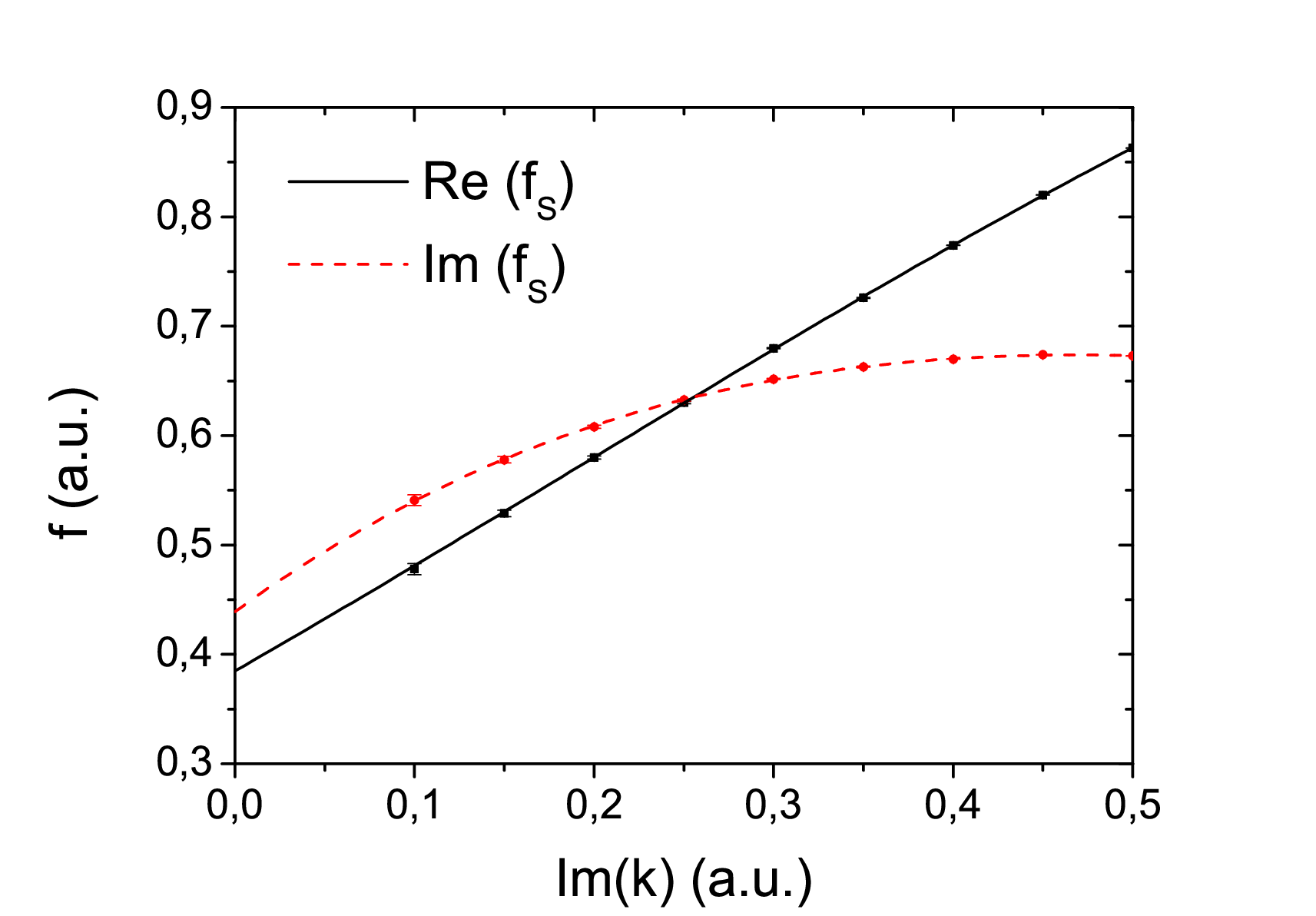,scale=0.5}
\begin{minipage}[t]{16.5 cm}
\caption{Electron scattering on hydrogen atom in its ground state
at $Re(k)=1.103$ a.u. Scattering amplitude dependence on the
imaginary part of the momenta. Calculated points by McDonald and
Nuttall~\cite{MDN69} are fitted using third order polynomial
function. \label{fig:eH_Nut}}
\end{minipage}
\end{center}
\end{figure}

Formally there is the liberty to choose the energy mapping $E'_c(E_c)$
between the incoming wave and scattered wave terms, as long as it
allows one to perform extrapolation to the real energy axis.
Obviously it should be a smooth function with the formal requirement
that the real energies are not affected by the mapping, i.e.
$E'_c(Re(E_c))=Re(E_c)$. Two natural choices exist for the mapping
function, namely $E'_c(E_c)=E_c$ and $E'_c(E_c)=Re(E_c)$. In the
pioneering work of McDonald and Nuttall~\cite{MDN69} the three-body
Coulomb problem was considered; therefore the second relation has
been chosen in order to avoid divergence of the inhomogeneous term
in eq.~(\ref{Schr_dr_nbodyt}). However the convergence of the
inhomogeneous term may be also provided by exponentially-bound
interactions. Based on our limited experience studying the 2-body
system we have found that the extrapolation procedure is more stable
using the $E'_c(E_c)=E_c$ mapping.

In Fig.~\ref{fig:eH_Nut} we present the behavior of the
spin-singlet amplitude as a function of the
imaginary part $Im(k)$ of the momentum for electron scattering on hydrogen atom in its ground state at $Re(k)=1.103$ a.u.. These calculations have
been performed
 by McDonald and Nuttall in their pioneering works~\cite{MDN69,mcdonald:71}
on the complex-energy method. At that time the numerical solution
of the full three-body problem was beyond the technical means and
the scattering amplitude was estimated to second order from the
variational principle of Schlessinger~~\cite{schlessinger:68}. We
may see a smooth amplitude dependence on the complex momentum,
thus enabling easy extrapolation to real momentum value. In
Fig.~\ref{fig:eH_Nut} a third order polynomial fit is used,
providing an extrapolated value $f_S=0.365(14)+0.440(14)i$ in full
agreement with the result of McDonald and Nuttall using rational
fraction fitting procedure.

The developments of  McDonald and Nuttall~\cite{MDN69,mcdonald:71}
based on the complex energy method have not been pursued in
configuration space. Nevertheless this method has been revived
recently by developments in momentum space. The momentum space
complex-energy method is presented in the next section.


%% file: Contrib_Arnas_cor.tex
\section{Complex energy method in momentum space \label{sec:cemm}}

As shown in previous sections, if one uses
differential equations in coordinate space for the
description of  few-particle scattering processes, one is faced
with nontrivial asymptotic boundary conditions when the channels
with three or more clusters become energetically open. In the
framework of momentum-space integral equations this gives rise
to integral kernels with a very complicated structure of
singularities. Furthermore, the complexity of the singularities
increases with the number of particles (likewise the complexity of the
wave function asymptotic form in configuration space). Formally,
this difficulty can be avoided by applying the complex energy method
of ref.~\cite{MDN69,mcdonald:71} described in a previous chapter to
momentum space calculations \cite{kamada:03a}.
The complex energy parameter used to damp outgoing waves in
configuration space calculations serves to smoothen  integral
kernel singularities present in momentum space calculations. Using
such a direct momentum-space approach the neutron-deuteron
scattering with simple model potentials was calculated in
refs.~\cite{kamada:03a,phyu:12}.

Nevertheless, only for sufficiently large complex energy parameter
the integral equation kernels become smooth enough to be integrated
without special care. On the other hand, in the three-particle
system the complex energy method seems to be unnecessary. Indeed,
the most sophisticated momentum-space calculations of
neutron-deuteron \cite{witala:01a,kuros:02b,deltuva:03a} and
proton-deuteron \cite{deltuva:05a,deltuva:05d,deltuva:09e} elastic
scattering and breakup and of three-body nuclear reactions
 \cite{deltuva:07d,deltuva:09b} are done directly
at real energies using  real-axis integration methods; thus, the
integral equation kernel singularities in the three-particle
system are well under control. In contrast, the only existing
momentum-space calculations for the scattering of four particles  above
 four-particle breakup threshold are done using the complex
energy method. First four-nucleon scattering calculations were
presented in ref. \cite{uzu:03a}; however, they employed  simple
separable potentials. Only very recently fully realistic
four-nucleon scattering calculations using modern nuclear
interactions and including the proton-proton Coulomb force have
been performed \cite{deltuva:12c,deltuva:13a,deltuva:13c}.
Important refinements of the complex energy method were developed
to improve its accuracy and practical applicability, given the
need to include a large number of partial waves in realistic
calculations. In this section the complex energy method with
emphasis on these special developments is summarized. The
four-nucleon system is employed to illustrate the method.

Four-nucleon scattering process may be described
exactly using
 the Alt, Grassberger, and Sandhas (AGS) equations \cite{grassberger:67}
for the symmetrized four-particle transition operators
$\mcu_{\beta \alpha}$ as derived in ref.~\cite{deltuva:07a},
where the nucleons are treated as identical particles in the isospin
formalism, i.e.,
\begin{subequations}  \label{eq:AGS}
\begin{align}
\mcu_{11}  = {}&   \eta (G_0 \, t \, G_0)^{-1}  P_{34} + \eta
P_{34} U_1 G_0 \, t \, G_0 \, \mcu_{11}
 + U_2   G_0 \, t \, G_0 \, \mcu_{21}, \label{eq:U11}  \\
\label{eq:U21} \mcu_{21} = {}&  (G_0 \, t \, G_0)^{-1}  (1 + \eta
P_{34})
+ (1 + \eta P_{34}) U_1 G_0 \, t \, G_0 \, \mcu_{11}, \\
\mcu_{12}  = {}&  (G_0 \, t \, G_0)^{-1} + \eta P_{34} U_1 G_0 \,
t \, G_0 \, \mcu_{12}  +
U_2   G_0 \, t \, G_0 \, \mcu_{22}, \label{eq:U12}  \\
\label{eq:U22} \mcu_{22} = {}& (1 + \eta P_{34}) U_1 G_0 \, t \,
G_0 \, \mcu_{12}.
\end{align}
\end{subequations}
Here, $ \eta = -1$ (+1) for identical fermions (bosons),
$\alpha=1$ corresponds to the $3+1$ partition (12,3)4 whereas
$\alpha=2$ corresponds to the $2+2$ partition (12)(34); there are
no other distinct  two-cluster partitions in the system of four
identical particles. The energy dependence of the operators arises
from the free resolvent
\begin{gather}\label{eq:G0}
G_0 = (Z - H_0)^{-1}
\end{gather}
with the complex energy parameter $Z = E+ i\varepsilon$ and the
free Hamiltonian $H_0$, while
\begin{gather} \label{eq:t}
t = v + v G_0 t
\end{gather}
 is the pair (12) transition matrix derived from the potential $v$, and
\begin{gather} \label{eq:AGSsub}
U_\alpha =  P_\alpha G_0^{-1} + P_\alpha t\, G_0 \, U_\alpha
\end{gather}
are the symmetrized 3+1 or 2+2 subsystem transition operators. For
the four-nucleon system the basis states are antisymmetric under
exchange of two particles in the subsystem (12), and, in the $2+2$
partition, also in the subsystem (34). The full antisymmetry is
ensured by the permutation operators $P_{ab}$ of particles $a$ and
$b$ with $P_1 =  P_{12}\, P_{23} + P_{13}\, P_{23}$ and $P_2 =
P_{13}\, P_{24}$.

The scattering amplitudes for two-cluster reactions at available
energy $E = \epsilon_\alpha + p_\alpha^2/2\mu_\alpha =
\epsilon_\beta + p_\beta^2/2\mu_\beta$ are obtained from the
on-shell matrix elements $  \langle \mbf{p}_{\beta}|
T_{\beta\alpha} |\mbf{p}_{\alpha} \rangle
  = S_{\beta\alpha}
\langle  \phi_{\beta} | \mcu_{\beta\alpha}| \phi_{\alpha} \rangle
$ in the limit $\varepsilon \to +0$. Here $|\phi_{\alpha} \rangle
$ is the Faddeev component of the asymptotic two-cluster state in
the channel $\alpha$, characterized by the bound state energy
$\epsilon_\alpha < 0$, the relative momentum $\mbf{p}_\alpha$, and
the reduced mass $\mu_\alpha$. Thus, depending on the isospin,
$\epsilon_1$
 is the ground state energy of $\He$ or $\Hh$, and
$\epsilon_2$ is twice the  deuteron  energy $\epsilon_d$.
$S_{\beta\alpha}$ are the symmetrization factors
\cite{deltuva:07a}, i.e., $S_{11} = 3$,
 $S_{12} = 2\sqrt{3}$, $S_{21} = \sqrt{3}$, and $S_{22} = 2$.
The amplitudes for breakup reactions are given by the integrals
involving $\mcu_{\beta\alpha}|\phi_{\alpha}\rangle$
\cite{deltuva:12a,deltuva:13a}, i.e.,
\begin{subequations} \label{eq:U0}
\begin{align}
 \langle \Phi_{3} |  T_{3 \alpha} | \Phi_{\alpha} \rangle
= {}&   S_{3\alpha} \langle \Phi_{3} | [(1 +\eta P_{34}) U_1 G_0
\, t \, G_0 \, \mcu_{1\alpha}  + U_2 G_0 \,  t \, G_0 \,
\mcu_{2\alpha} ]
| \phi_{\alpha} \rangle, \\
\langle \Phi_{4} |  T_{4 \alpha} | \Phi_{\alpha} \rangle = {}&
S_{4\alpha}  \langle \Phi_{4} | (1+P_1) \{ [1+\eta P_{34}(1+P_1)]
 t \, G_0    U_1 G_0 \, t \, G_0 \, \mcu_{1\alpha} 
+ (1+P_2) t \, G_0    U_2 G_0 \,  t \, G_0 \, \mcu_{2\alpha} \} |
\phi_{\alpha} \rangle
\end{align}
\end{subequations}
for three- and four-cluster breakup, respectively. The
symmetrization factors are $S_{31} = \sqrt{3}$,
 $S_{32} = 2$, $S_{41} = \sqrt{3}$, and $S_{42} = 2$
where the asymptotic three- and four-cluster channel states $|
\Phi_{3} \rangle$ and $ | \Phi_{4} \rangle $ are antisymmetrized
(symmetrized for bosons) with respect to the pair (12).

The AGS equations \eqref{eq:AGS} are solved in the momentum-space
partial-wave framework. Two different types of basis states $|k_x
k_y k_z \nu \rangle_\alpha $ with $\alpha = 1$ and 2 are employed.
All discrete quantum numbers are abbreviated by $\nu$, while
 $k_x$, $k_y$, and $k_z$ denote  magnitudes of the Jacobi momenta.
For $\alpha=1$ the Jacobi momenta describe the relative motion in
the 1+1, 2+1, and 3+1 subsystems and are expressed in terms of
single particle momenta $\mbf{k}_{a}$ as
\begin{subequations} \label{eq:jacobi1}
\begin{align}
\mbf{k}_x = {}& \frac12(\mbf{k}_2 -\mbf{k}_1), \\
\mbf{k}_y = {}& \frac13[2\mbf{k}_3 -(\mbf{k}_1+\mbf{k}_2)], \\
\mbf{k}_z = {}& \frac14[3\mbf{k}_4
-(\mbf{k}_1+\mbf{k}_2+\mbf{k}_3)],
\end{align}
\end{subequations}
while for $\alpha=2$ they describe the relative motion in the 1+1,
1+1, and 2+2 subsystems, i.e.,
\begin{subequations} \label{eq:jacobi2}
\begin{align}
\mbf{k}_x = {}& \frac12(\mbf{k}_2 -\mbf{k}_1), \\
\mbf{k}_y = {}& \frac12(\mbf{k}_4 -\mbf{k}_3), \\
\mbf{k}_z = {}& \frac12[(\mbf{k}_4+\mbf{k}_3)
-(\mbf{k}_1+\mbf{k}_2)].
\end{align}
\end{subequations}
The reduced masses associated with Jacobi momenta $k_x$ and $k_y$
in the partition $\alpha$ will be denoted by $\mu_{\alpha x}$ and
$\mu_{\alpha y}$, respectively.

An explicit form of integral equations is obtained by inserting
the respective completeness relations
\begin{equation} \label{eq:k1k}
1 = \sum_{\nu} \int_0^\infty |k_x k_y k_z \nu \rangle_\alpha k_x^2
dk_x \, k_y^2 dk_y \, k_z^2 dk_z \, {}_{\alpha}\langle k_x k_y k_z
\nu |
\end{equation}
between all operators in Eqs.~\eqref{eq:AGS}. The integrals are
discretized using Gaussian quadrature rules \cite{press:89a}
turning  Eqs.~\eqref{eq:AGS} into a system of linear equations as
described in ref.~\cite{deltuva:07a}. However, in the limit
$\varepsilon \to +0$ needed for the calculation of the observables
the kernel of the AGS equations contains integrable singularities.
At $E + i\varepsilon - \epsilon_{\alpha} - k_z^2/2\mu_\alpha  \to
0 $ the subsystem transition operator in the bound state channel
has the pole
\begin{gather} \label{eq:Bpole}
G_0 U_\alpha G_0  \to \frac{P_\alpha |\phi_\alpha \rangle
S_{\alpha\alpha} \langle \phi_\alpha | P_\alpha} {E + i\varepsilon
- \epsilon_{\alpha} - k_z^2/2\mu_\alpha} .
\end{gather}
Furthermore, at $E + i\varepsilon - \epsilon_{d} -
k_y^2/2\mu_{\alpha y} -k_z^2/2\mu_\alpha \to 0$ the two-nucleon
transition matrix in the channel with the deuteron quantum numbers
for the pair (12) has the pole
\begin{gather} \label{eq:tpole}
 t \to  \frac{v |\phi_d \rangle \langle \phi_d | v }
{E + i\varepsilon - \epsilon_{d} - k_y^2/2\mu_{\alpha y} -
k_z^2/2\mu_\alpha},
\end{gather}
with $|\phi_d \rangle$ being the pair (12) deuteron wave function.
Finally, the free resolvent \eqref{eq:G0} obviously becomes
singular at $E + i\varepsilon - k_x^2/2\mu_{\alpha x} -
k_y^2/2\mu_{\alpha y} - k_z^2/2\mu_\alpha \to 0$.

At energies below the three-cluster threshold only
singularities of the type \eqref{eq:Bpole} are present. In previous
momentum-space calculations \cite{deltuva:07a} they were treated
reliably by the subtraction technique. However, above the
four-body breakup threshold all three kinds of
 singularities are present. Their interplay with permutation operators
and basis transformations leads to a very complicated singularity
structure of the AGS equations.

This  difficulty can be formally avoided by following the ideas
proposed in
Refs.~\cite{schlessinger:68,MDN69,mcdonald:71,kamada:03a}, i.e.,
by performing calculations for a set of finite $\varepsilon
> 0$ values where the kernel contains no singularities and then
extrapolating the results to the $\varepsilon \to +0$ limit. The
extrapolation is usually done using the point method
\cite{schlessinger:68}: The scattering amplitudes $T(Z_n)$ (for
brevity the dependence on the momenta and channels is suppressed)
are calculated for the set of complex energy values $\{Z_n\}$, $n
= 1,2,\ldots,N+1$, and then the amplitudes at a desired $Z$, i.e.,
$\varepsilon \to +0$,
 are obtained using analytic continuation via continued fraction
\begin{gather} \label{eq:cfr}
T_N(Z) = \frac{T(Z_1)}{1+} \, \frac{a_1(Z-Z_1)}{1+} \,
\frac{a_2(Z-Z_2)}{1+} \ldots \frac{a_N(Z-Z_N)}{1}.
\end{gather}
Demanding that $T_N(Z_{n+1}) = T(Z_{n+1})$ the expansion
coefficients $a_n$ are obtained recursively as
\begin{gather} \label{eq:cfra}
a_n = \frac{1}{Z_n-Z_{n+1}} \left[ 1+
\frac{a_{n-1}(Z_{n+1}-Z_{n-1})}{1+} \,
\frac{a_{n-2}(Z_{n+1}-Z_{n-2})}{1+} \ldots
\frac{a_{1}(Z_{n+1}-Z_{1})}{1 - T(Z_1)/T(Z_{n+1})} \right]
\end{gather}
starting with $a_1 = [T(Z_1)/T(Z_2)-1]/(Z_2-Z_1)$.

However, this extrapolation method as well as alternative choices
are only precise for not too large $\varepsilon$ values. On the
other hand, for small $\varepsilon$ the kernel of the AGS
equations, although formally being nonsingular, may exhibit a
quasi-singular behavior thereby requiring dense grids for the
numerical integration. This is no problem in simple model
calculations with rank-one separable potentials and very few
channels \cite{uzu:03a} where one can use a large number of grid
points. However, in practical calculations with realistic
potentials and large number of partial waves necessary for the
convergence one has to keep the number of integration grid points
as small as possible  and therefore a more sophisticated integration
method is needed.

An important technical improvement when calculating
$\mcu_{\beta\alpha}$ at finite $\varepsilon $ was introduced in
ref.~\cite{deltuva:12c}. The method of special weights for
numerical integrations involving any of the above-mentioned
quasi-singularities is used, i.e.,
\begin{gather} \label{eq:wspc}
\int_{a}^{b} \frac{f(x)}{x_0^n + iy_0 - x^n} dx \approx
\sum_{j=1}^{N} f(x_j) w_j(n,x_0,y_0,a,b).
\end{gather}
The quasi-singular factor $(x_0^n + iy_0 - x^n)^{-1}$ is separated
and absorbed into the special integration weights
$w_j(n,x_0,y_0,a,b)$. The set of $N$ grid points $\{x_j\}$ where
the remaining  smooth function $f(x)$ has to be evaluated is
chosen the same as for the standard Gaussian quadrature. However,
while the standard weights are real \cite{press:89a}, the special
ones $w_j(n,x_0,y_0,a,b)$ are complex. They are chosen such that
for a set of $N$ test functions $f_j(x)$ the result
\eqref{eq:wspc} is exact. A convenient and reliable choice of
$\{f_j(x)\}$ are the $N$ spline functions  $\{S_j(x)\}$ referring
to the grid $\{x_j\}$; their construction and properties are
described in Refs.~\cite{press:89a,boor:78a,gloeckle:82a}. The
corresponding special weights are
\begin{gather} \label{eq:wss}
 w_j(n,x_0,y_0,a,b) =
\int_{a}^{b} \frac{S_j(x)}{x_0^n + iy_0 - x^n} dx,
\end{gather}
where the integration can be performed either analytically or
numerically using a sufficiently dense grid. This choice of special
weights guarantees accurate results for quasi-singular integrals
\eqref{eq:wspc} with any $f(x)$ that can be accurately
approximated by the spline functions  $\{S_j(x)\}$.

In the integrals over the momentum variables one has
 $n=2$,  $a=0$, and $b\to \infty$. For example,
when solving the Lippmann-Schwinger equation \eqref{eq:t} the
integration variable in eq.~\eqref{eq:wspc} is the momentum $k_x$
with
 $x_0^2 = 2\mu_{\alpha x}(E - k_y^2/2\mu_{\alpha y} -k_z^2/2\mu_\alpha)$
and $y_0 = 2\mu_{\alpha x}\varepsilon $. Alternatively, the
quasi-singularity can be isolated in a narrower interval $0 < a <
b < \infty$ and treated by special weights only there.

 Other numerical techniques for solving the four-nucleon AGS equations
are taken over from ref.~\cite{deltuva:07a}. They include Pad\'{e}
summation \cite{baker:75a} of Neumann series for the transition
operators $U_\alpha$ and $\mcu_{\beta \alpha}$ using the algorithm
of  ref.~\cite{chmielewski:03a} and the treatment of permutation
operators (basis transformations) using the spline interpolation.
The specific form of the permutation operators \cite{deltuva:07a}
leads to a second kind of quasi-singular integrals \eqref{eq:wspc}
with  $n=1$, $a=-1$, $b=1$, where the integration variable
 $x =  \hat{\mbf{k}}'_y \cdot \hat{\mbf{k}}_y$
or  $\hat{\mbf{k}}'_z \cdot \hat{\mbf{k}}_z$ is the cosine of
the angle between the respective initial and final momenta.

The above integration method is not sufficient in the vanishing
$\varepsilon $ limit since for $n=1$ and $y_0=0$ the result of the
integral \eqref{eq:wspc} contains the contribution $f(x_0)
\ln[(x_0+1)/(x_0-1)]$ with logarithmic singularities at $x_0 = \pm
1$. At finite small  $\varepsilon $  the result of
\eqref{eq:wspc} may exhibit a quasi-singular behavior. However,
since the logarithmic  quasi-singularity is considerably weaker
than the pole quasi-singularity, for not too small  $\varepsilon $
it is sufficient to use the standard integration.

Below the three-cluster breakup threshold direct calculations at
real energies using the subtraction technique \cite{deltuva:07a}
for the treatment of the bound state poles \eqref{eq:Bpole} are
available. Comparison with these results proves the extreme
accuracy of the complex energy method with special integration
weights. On the other hand, in this regime the real energy method
\cite{deltuva:07a} is much more efficient as it does not require
extrapolation and single $\varepsilon = 0$ calculation suffice.


\begin{table}[!]
\begin{tabular}{l|*{10}{c}}
\hline $[\varepsilon_{\mathrm{min}},\varepsilon_{\mathrm{max}}]$ &
$\delta ({}^1S_0)$ & $\eta ({}^1S_0)$  & $\delta ({}^3S_1)$ &
$\eta ({}^3S_1)$  & $\delta ({}^3D_1)$ & $\eta ({}^3D_1)$  &
$\delta ({}^3P_0)$ & $\eta ({}^3P_0)$  & $\delta ({}^3P_2)$ &
$\eta ({}^3P_2)$
\\  \hline
$[1.0,2.0]$ & 62.63 & 0.990 & 72.87 & 0.983 & 3.39 & 0.933 & 43.03 & 0.959 & 65.27 & 0.950 \\
$[1.2,2.0]$ & 62.60 & 0.991 & 72.88 & 0.982 & 3.40 & 0.933 & 43.04 & 0.959 & 65.29 & 0.951 \\
$[1.4,2.0]$ & 62.67 & 0.991 & 72.93 & 0.983 & 3.39 & 0.933 & 43.03 & 0.958 & 65.27 & 0.950 \\
$[1.2,1.8]$ & 62.65 & 0.992 & 72.97 & 0.983 & 3.39 & 0.933 & 43.03 & 0.959 & 65.28 & 0.950 \\
1.4 &         73.37 & 0.916 & 83.93 & 0.978 & 3.80 & 0.929 & 44.77 & 0.840 & 67.38 & 0.933 \\
\hline
\end{tabular}
\caption{ \label{tab:conv} Elastic phase shifts (in degrees) and
inelasticities in selected partial waves for $n$-$\Hh$ scattering
at 22.1 MeV neutron energy. Results for INOY04 potential obtained
using different sets of $\varepsilon$ values  ranging from
$\varepsilon_{\mathrm{min}}$ to $\varepsilon_{\mathrm{max}}$ (in
MeV) are compared. In the last line the predictions with
$\varepsilon = 1.4$ MeV without extrapolation are given.}
\end{table}

However, above the three- and four-cluster  breakup threshold the
real-energy technique becomes extremely complicated and has not been
implemented. Therefore the only existing momentum-space
calculations are performed using the complex energy method whose
numerical reliability at not too high energies is demonstrated in
ref.~\cite{deltuva:12c}. The test uses realistic dynamics,
namely, the high-precision inside-nonlocal outside-Yukawa (INOY04)
two-nucleon potential  by Doleschall
\cite{doleschall:04a,lazauskas:04a} and includes a large number of
four-nucleon partial waves sufficient for the convergence. The
chosen potential nearly reproduces experimental binding energies
of $\Hh$ (8.48 MeV) and $\He$ (7.72 MeV) without an irreducible
three-nucleon force. There are too many numerical parameters
(numbers of points for various integration grids) to demonstrate
the stability of the calculations with respect to each of them
separately. It was found that 10 grid points are sufficient for
all angular integrations but 30 to 40 grid points are needed for
the discretization of Jacobi momenta. This is more than 20 to 25
grid points needed for the real or complex energy calculations
below the three-cluster breakup threshold. The $\varepsilon \to
+0$ extrapolation yields stable results only if sufficiently small
$\varepsilon$ are considered and at each of them the respective
calculations are numerically well converged. This is achieved as
Table \ref{tab:conv} demonstrates. It collects results for phase
shifts $\delta$ and inelasticities $\eta$ for $n$-$\Hh$ scattering
at $E_n = 22.1$ MeV neutron energy obtained via
$\varepsilon \to +0$ extrapolation using different $\varepsilon$
sets ranging from $\varepsilon_{\mathrm{min}}$ to
$\varepsilon_{\mathrm{max}}$ with a step of 0.2 MeV. One finds a
very good agreement between the results obtained with
$[\varepsilon_{\mathrm{min}},\varepsilon_{\mathrm{max}}] =$
[1.0,2.0], [1.2,2.0], [1.4,2.0], and [1.2,1.8] MeV, confirming the
reliability of the calculations. In addition, we also list the
predictions referring to $\varepsilon = 1.4$ MeV  without
extrapolation that don't have any physical meaning. The difference
between
 $\varepsilon \to +0$ and $\varepsilon = 1.4$ MeV results
demonstrates the importance of the extrapolation. The stability of
the results with respect to changes in
$[\varepsilon_{\mathrm{min}},\varepsilon_{\mathrm{max}}]$ is very
good. The variations are slightly larger in  the $S$ waves where
also the difference between the finite  $\varepsilon$ and
$\varepsilon \to +0$ results is most sizable.

\begin{figure}[!]
\begin{center}
\includegraphics[scale=0.69]{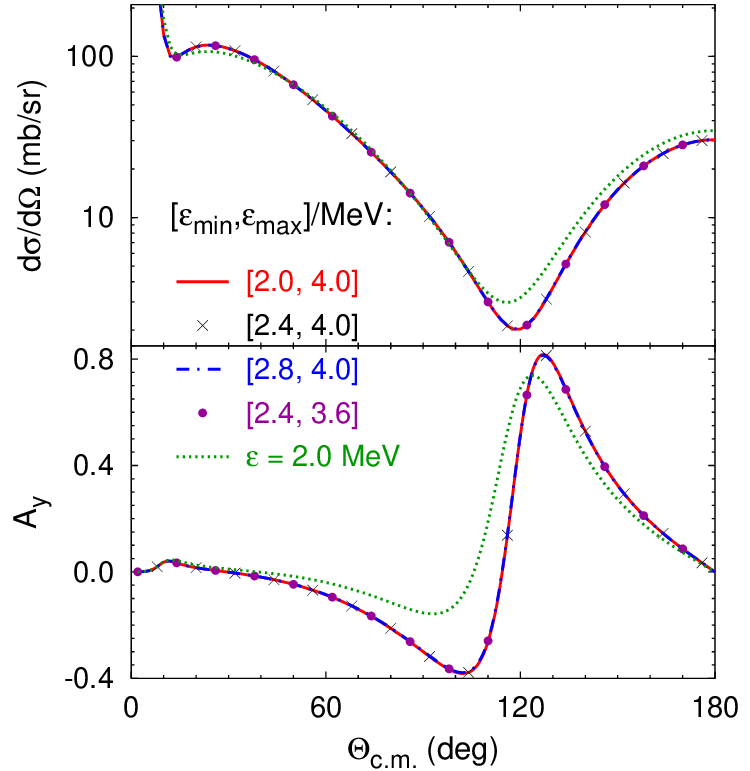}
\end{center} 
\caption{ \label{fig:conv} (Color online) Differential cross
section and proton analyzing power for elastic $p$-$\He$
scattering at 25 MeV proton energy as functions of c.m. scattering
angle. Results obtained using different sets of $\varepsilon$
values ranging from $\varepsilon_{\mathrm{min}}$ to
$\varepsilon_{\mathrm{max}}$ with the step of 0.4 MeV are
compared; they are indistinguishable. The dotted curves refer to
the  $\varepsilon = 2.0$ MeV calculations without extrapolation
that have no physical meaning but demonstrate the importance of
the extrapolation. }
\end{figure}

Another example for the stability of the $\varepsilon \to +0$
extrapolation is presented Fig.~\ref{fig:conv} where the
differential cross section $d\sigma/d\Omega$ and proton analyzing
power $A_y$ for elastic $p$-$\He$ scattering at $E_p = 25$ MeV
proton energy are shown. Again, the stability of the results with
respect to changes in
$[\varepsilon_{\mathrm{min}},\varepsilon_{\mathrm{max}}]$ is very
good. One finds a very good agreement between the results obtained
with $[\varepsilon_{\mathrm{min}},\varepsilon_{\mathrm{max}}] =$
[2.0,4.0], [2.4,4.0], [2.8,4.0], and [2.4,3.6] MeV, confirming the
reliability of the employed method. From  Table \ref{tab:conv}
and Fig.~\ref{fig:conv} one can conclude that with a proper
$\varepsilon$ choice
 as few as four different $\varepsilon$ values are sufficient to obtain
the physical $\varepsilon \to +0$ results with good accuracy.

As pointed out in ref.~\cite{deltuva:12c},
 the calculations keeping the same grids but
with standard integration weights fail completely at $\varepsilon$
values from Table \ref{tab:conv}, with the errors of the
$\varepsilon \to +0$ extrapolation being up to 10 \% for phase
shifts and up to 25 \% for inelasticity parameters. On the other
hand, at large $\varepsilon > 4$ MeV the two integration methods
agree  well but the $\varepsilon \to +0$  extrapolation has at
least one order of magnitude larger inaccuracies than those
presented in Table \ref{tab:conv}.

\begin{figure}[!]
\begin{center}
\includegraphics[scale=0.86]{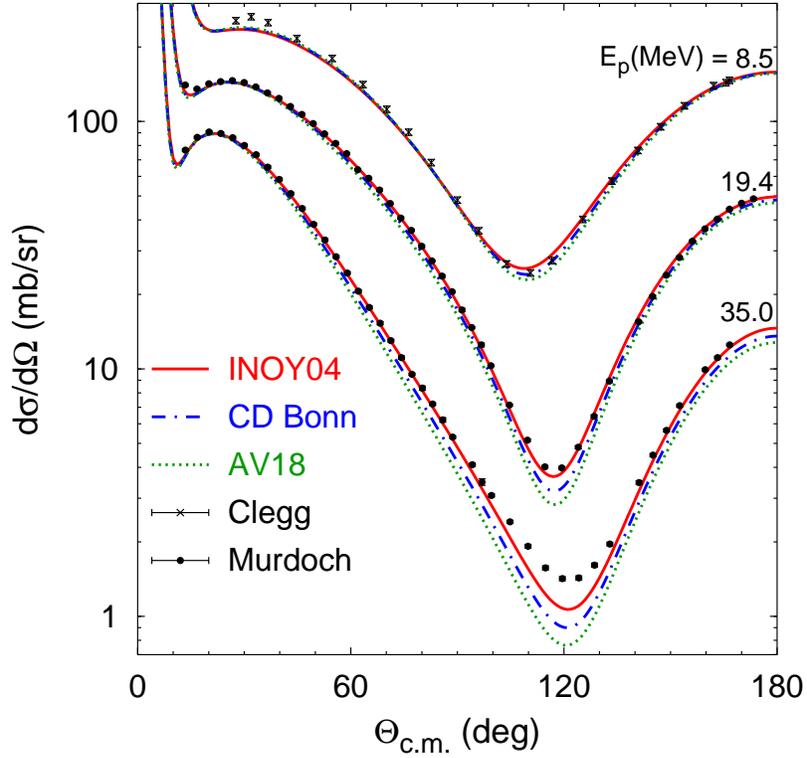}
\end{center} 
\caption{ \label{fig:dcs1} (Color online) Differential cross
section for elastic $\pHe$ scattering at 8.52, 19.4, and 35.0 MeV
proton energy as function of the c.m. scattering angle. Results
obtained with INOY04 (solid curves), CD Bonn (dashed-dotted
curves), and AV18 (dotted curves) potentials are compared with the
experimental data from Refs.~\cite{clegg:64,murdoch:84a}.}
\end{figure}

An example for the physics results obtained with various realistic
high-precision NN potentials, namely, the Argonne (AV18) potential
\cite{wiringa:95a}, the charge-dependent Bonn potential (CD Bonn)
\cite{machleidt:01a}, and the INOY04 potential, is presented in
Fig.~\ref{fig:dcs1}. The differential cross section
$d\sigma/d\Omega$ for elastic $\pHe$ scattering at a number of
proton energies ranging from $E_p = 8.5$ to 35.0 MeV is shown.
This observable decreases rapidly with the increasing energy and
also changes the shape; the calculations describe the energy and
angular dependence of the experimental data fairly well. Below
$E_p = 10$ MeV the  experimental data are slightly underpredicted
at forward angles as happens also at energies below the
three-cluster breakup threshold \cite{viviani:11a,deltuva:07b}. At
the minimum the $d\sigma/d\Omega$ predictions scale with the $\He$
binding energy:  the weaker the $\He$  binding the lower the dip
of $d\sigma/d\Omega$ that is located between $\Theta_{\cm} =
105^{\circ}$ and $\Theta_{\cm} = 125^{\circ}$. The scaling is more
pronounced at higher energies.
 For the INOY04 potential that fits the $\He$ binding energy,
one gets excellent agreement in the whole angular region up to
$E_p\simeq 20$ MeV but, as the energy increases, the calculated
cross section starts underpredicting the data. This may be a sign
for the need of a three-nucleon force. More detailed study of
the elastic $\pHe$ scattering, including various spin observables
like analyzing powers, spin-correlation and spin-transfer
coefficients, can be found in ref.~\cite{deltuva:13c}.

In summary, realistic and fully converged four-nucleon scattering
calculations above the four-nucleon breakup threshold becomes
feasible using the complex energy method with a special integration
technique in the momentum-space framework.
The only stumbling block at the present time for momentum-space
four-nucleon calculations is adding a state of the art static
three-nucleon force to the underlying realistic NN force.
However, effective 3N and 4N forces have been included
via explicit NN-N$\Delta$ coupling in the two-baryon potential,
both below \cite{deltuva:08a} and above \cite{deltuva:13c}
breakup threshold.  Extension of the method to other reactions in the
4N system is in progress.

\newpage

\section{Momentum lattice technique \label{sec:mml}}

The momentum lattice technique developed in
Refs.~\cite{kukulin:07a,pomerantsev:09a,rubtsova:09a,rubtsova:12a}
is based on the idea of discretization of all momentum variables
using finite wave-packet basis of the so-called $L_2$ type. In
this respect it has some similarity with the continuum discretized
coupled channels (CDCC) method \cite{austern:87}. In contrast to
free waves employed in the standard momentum-space scattering
calculations, $L_2$ basis states are square-integrable functions
much like the bound-state wave functions. Thus, in this approach
the few-body scattering problem is  formulated in a Hilbert space
of few-body {normalized} states, and all involved operators are
approximated by finite-dimensional matrices. In this respect it is
similar to the bound-state problem. The method has been
successfully applied to study neutron-deuteron elastic scattering
and breakup using semirealistic as well as  realistic
interactions~\cite{rubtsova:09a,rubtsova:12a,kukulin:12a}. In this
section these developments are summarized.

The description of three identical particles uses the standard
Jacobi momenta $\mbf{p}$ and  $\mbf{q}$ that coincide with
$\mbf{k}_x$  and $\mbf{k}_y$ in eq.~(\ref{eq:jacobi1}),
respectively. The corresponding continuum partial-wave states are
normalized to Dirac $\delta$-functions (note, however, a different
convention as compared to previous section) $\langle
p'|p\rangle=\delta(p'-p)$  and $\langle q'|q\rangle=\delta(q'-q)$
where the dependence on the angular momentum, spin, and isospin
quantum numbers is suppressed for simplicity. The continuum part of the
momenta $0\le p \le p_{\rm max}$ is divided into $M$
nonoverlapping bins $\MD_i \equiv [p_{i-1},p_i]$ with
$i=1,\ldots,M$
 whereas the high-momentum part of the spectrum above  $p_{\rm max}$ is
neglected. In the same manner the continuum part of momenta $0\le
q \le q_{\rm max}$ is divided into $N$ nonoverlapping bins $\BMD_j
\equiv [q_{j-1},q_j]$ and the high-momentum part above  $q_{\rm
max}$ is again neglected. The
 widths of the momentum bins are $d_i=p_i-p_{i-1}$ and
$\bar{d}_j=q_j-q_{j-1}$,  respectively. The partial-wave packet
$L_2$ basis is constructed as
\begin{eqnarray}
\label{ip}
|\mathfrak{p}_i\rangle=\frac{1}{\sqrt{A_i}}\int_{\MD_i}dp
f(p)|p\rangle,\quad i=1,\ldots,M,\\
\label{iq}
|\mathfrak{q}_j\rangle=\frac{1}{\sqrt{B_j}}\int_{\BMD_j}dq
w(q)|q\rangle,\quad j=1,\ldots,N.
\end{eqnarray}
Here $f(p)$ and $w(q)$ are freely chosen weight functions with the
corresponding normalization factors
\begin{eqnarray}
\label{norm} A_i=\int_{\MD_{i}}dp |f(p)|^2, \\
B_j=\int_{\BMD_j}dq|w(q)|^2
\end{eqnarray}
ensuring the desired normalization
$\langle\mathfrak{p}_{i'}|\mathfrak{p}_{i}\rangle=\delta_{i'i} $
and
$\langle\mathfrak{q}_{j'}|\mathfrak{q}_{j}\rangle=\delta_{j'j}$ of
the wave packet states. One of the simplest possible choices for
the  weight functions is  $f(p)= 1$ resulting in $A_i=d_i$. In
this case the momentum representation  of the wave packet
 \begin{equation}
 \label{proj_rule}
 \langle p|\mathfrak{p}_i\rangle=\frac{\vartheta(p\in \MD_i)}{\sqrt{d_i}},
 \end{equation}
takes a form of step-like function where $\vartheta(p\in \MD_i)
\equiv \vartheta(p-p_{i-1}) \vartheta(p_i-p)$.

The three-body wave packet states  are built as direct products of
the above wave packets for the pair and spectator particle motion,
i.e., $|\mathfrak{p}_i\rangle\otimes|\mathfrak{q}_j\rangle$. Since
the basis functions are the products of both step-like functions
in variables $p$ and $q$,  the solution of the three-body
scattering problem in such a basis corresponds to a formulation of
the scattering problem   on a two-dimensional momentum lattice,
with lattice cells $\MD_{ij}=\MD_i\otimes\BMD_j$. Using such a
lattice basis, in principle one could solve the three-body
scattering problem by projecting  all the scattering operators
onto wave packet states. In this representation the free
Hamiltonian $H_0$ as well as the free resolvent $G_0 =
(Z-H_0)^{-1}$ remain diagonal; their matrix elements have explicit
analytical forms \cite{kukulin:07a}. Other operators, e.g., the
three-body transition operator ${U}_1$ of eq.~(\ref{eq:AGSsub}),
 must be transformed into lattice basis
according to relations (\ref{ip}) and  (\ref{iq}) thereby becoming
finite-dimensional matrices
\begin{equation} \label{op}
 [{\mathbb U}_1]_{i'j',ij} =
\frac{1}{\sqrt{A_{i'}B_{j'}A_{i}B_{j}}}
\int_{\MD_{i'j'}}dp'dq'\int_{\MD_{ij}}dpdq f(p')w(q') \langle
p'q'|U_1|pq\rangle f(p)w(q).
\end{equation}

As an alternative to the free wave packets discussed so far, one
 may consider scattering wave packets
for the correlated pair of particles constructed as
\begin{equation}
\label{ipsct}
|\psi(\mathfrak{p}_i)\rangle=\frac{1}{\sqrt{A_i}}\int_{\MD_i}dp
f(p)|\psi(p)\rangle.
\end{equation}
Here $|\psi(p)\rangle$ is the exact scattering wave function
corresponding to the channel Hamiltonian $H_1 = H_0+v_1$ with
$v_1$ being the potential for the pair 1 consisting of particles 2
and 3 in the odd-man-out notation. Of course, the binning $\MD_i$,
the weight function $f(p)$, and the respective normalization
factors $A_i$ can be chosen differently as compared to
eq.~(\ref{ip}). If the  Hamiltonian $H_1$ supports also a bound
state, the set of scattering wave packets (\ref{ipsct}) has to be
accomplished with this bound state wave function $|\psi_0\rangle$
to form an orthonormalized basis. In such basis the  channel
Hamiltonian $H_1$ and the respective channel resolvent $G_1 =
(Z-H_1)^{-1}$ are diagonal \cite{kukulin:07a}. This suggests an
alternative form of three-body scattering equations instead of
(\ref{eq:AGSsub}) for the transition operator $U_1$. Given the
identity $t G_0 = v_1 G_1$ and that on-shell
$(G_0^{-1}-v_1)|\psi_0\rangle = 0$, the  three-body transition
operator $\tilde{U}$ satisfying the integral equation
\begin{equation}
\label{pvg} \tilde{U}=P_1v_1+P_1v_1G_1 \tilde{U},
\end{equation}
is equivalent to $U_1$ on- and half-shell and therefore describes
the same scattering process.
An essential advantage of this approach is that the singularities
are integrated out when calculating the matrix elements of the operators
in eq.~(\ref{pvg}) in the momentum-lattice basis such that the resulting
matrix equations are nonsingular. Furthermore, explicit calculation
of the two particle transition matrix $t$ and its interpolations
are avoided.

In practical calculations the knowledge of the exact scattering
wave packets (\ref{ipsct}) is not necessary since they are
approximated by the  pseudostates
$|\tilde{\psi}(\mathfrak{p}_i)\rangle$ obtained by the
diagonalization of the channel Hamiltonian $H_1$ in the basis of
free  wave packets (\ref{ip}), i.e.,
\begin{equation}
  \label{exp_z}
  |\tilde{\psi}(\mathfrak{p}_k)\rangle=\sum_{i=1}^M
O_{ki}|\mathfrak{p}_i\rangle,\quad    k=1,\ldots,M,
  \end{equation}
with $O_{ki}$ being the elements of the respective transformation
matrix. It has been demonstrated in ref.~\cite{kukulin:07a} that
the properties of $|{\psi}(\mathfrak{p}_i)\rangle$ and
$|\tilde{\psi}(\mathfrak{p}_i)\rangle$
 are quite similar thereby justifying the above approximation.
To solve the scattering equations in the pseudostate basis
(\ref{exp_z}) the corresponding transformations for the matrices
of all involved operators has to be done. Finally, since the basis
states are step-like functions of momenta and/or energies, the
energy-averaging procedure has to be applied to obtain the breakup
amplitudes from the solutions of the finite-dimensional matrix
equations for $\tilde{\mathbb U}$, as described in detail in
ref.~\cite{rubtsova:12a}.

The above momentum lattice method has been first tested in a model
study of neutron-deuteron elastic scattering and breakup with
semirealistic interactions limited to $S$
waves~\cite{rubtsova:09a,rubtsova:12a}. More recently the results
for elastic neutron-deuteron scattering were   obtained also with
realistic interactions~\cite{kukulin:12a}.  The method was able to
reproduce reasonably well the results obtained in standard
momentum-space calculations, but, in its presently available
technical implementation~\cite{kukulin:12a} is still less
efficient than the standard momentum-space methods of
refs.~\cite{gloeckle:96a,deltuva:03a}\footnote{For example, the
number of bins had to be at least 200 for each Jacobi momentum,
whereas the standard momentum-space methods achieve high accuracy
typically using 30 to 40 grid points~\cite{deltuva:03a}. Thus its
technical realization roughly requires to perform by two orders of
magnitude more CPU operations.}.
Thus, it remains yet unclear if the momentum lattice method could
be extended beyond the A=3 case in realistic few-body scattering
calculations. On the other hand,  the momentum lattice method has
additional advantages when performing calculations at several
energies simultaneously, and in systems with charged particles
where one could use Coulomb wave packets thereby avoiding
screening and renormalization procedure. Due to the finite $L_2$
character of the wave packet basis which is rather similar to the
harmonic oscillator basis one can use the Hamiltonian
diagonalisation procedure to find the scattering states and the
S-matrix, thereby avoiding the need for the solution of scattering
equations at all \cite{pomerantsev:pc}. Finally, the momentum
lattice method is better suited for implementing the calculations
on graphic processor units, leading to significant gains in speed
\cite{kukulin:12a}.

We would like to mention the existence of other techniques to compute
scattering observables which are also based on bound state
solutions in a discretized space, although in a quite a different spirit of
what has been presented above.
They were developed by Luscher and collaborators \cite{Luscher} in a series
of papers devoted to Lattice QCD \cite{Les_Houches_2009}.
These techniques are based on computing the volume dependence of
the confined solutions and obtaining from them the corresponding low energy
scattering parameters.
First formulated in the framework of non relativistic quantum mechanics,
they were aimed to extract the scattering observables from the solution of
a Quantum Field theoretical problem
obtained using the Feynman path integral formulation of the theory in an
euclidean discretized
space-time with periodic and/or anti-periodic boundary conditions.
They overcome this way the no-go theorem of Maiani and Testa
\cite{MT_NPB_245_90}, damming the access to the scattering
observables from any euclidean discretized version of a Quantum Field
theory in a finite volume.
The present applications of this technique are however limited
to the scattering of simple two-body composite systems,
far from the complexity of the asymptotic many-body multichannel
wavefunctions, esspecially in presence of the breakup channels.
The interested reader can take benefit in consulting the recent reviews
on this topic in different fields of application
\cite{Lee_lattice_2009,Beane_LQCD_2011}.